\newif\ifsinglecol
\singlecoltrue 

\ifsinglecol
\documentclass[11pt,draftcls, onecolumn]{IEEEtran}
\else
\documentclass[10pt,final, twocolumn]{IEEEtran}
\fi 

\usepackage{times}
\usepackage{amsmath,dsfont}
\usepackage{amssymb,amsthm}
\usepackage{epsfig,verbatim}
\usepackage{subfigure}
\usepackage{setspace}
\usepackage{color}
\usepackage{cite}
\usepackage{epstopdf}
\usepackage{graphics}
\usepackage{enumitem}
\usepackage{accents}
\usepackage{acronym}
\usepackage{enumitem}
\usepackage[bookmarks,colorlinks]{hyperref}
\usepackage{booktabs}
\usepackage{tabularx} 
\usepackage{adjustbox}

\newtheorem{definition}{Definition}
\newtheorem{theorem}{Theorem}
\newtheorem{corollary}{Corollary}
\newtheorem{proposition}{Proposition}
\newtheorem{lemma}{Lemma}
\newtheorem{remark}{Comment}

\newcommand{\E}{\mathds{E}}

\newcommand{\utilde}{\underaccent{\tilde}}

\newcommand{\Rx}{{\bf R}_{\bf x}} 
\newcommand{\Cx}{{\bf C}_{\bf x}}
\newcommand{\cx}{{\bf c}_{\bf x}}

\newcommand{\RxProd}[2]{{\bf L}_{#1,#2}^{\tilde{\bf R}}} 
\newcommand{\FProd}[2]{{\bf L}_{#1,#2}^{\bf F}} 
\newcommand{\bk}[1]{{\bf f}_{#1}}
\newcommand{\pk}[1]{{\bf z}_{#1}}
\newcommand{\sk}[1]{{\bf s}_{#1}}

\newcommand{\ho}{{\bf h}_{{\rm TA}}} 
\newcommand{\he}{{\bf{\bar h}}}

\newcommand{\ERx}{\tilde{\bf R}_{\bf x}}

\newcommand{\SigV}{\sigma_{v}^2}

\newcommand{\F}{{\bf F}}
\newcommand{\A}{{\bf A}} 
\newcommand{\B}{{\bf B}} 
\newcommand{\Hmat}{{\bf H}} 
\newcommand{\Pmat}{{\bf P}} 
 
\newcommand{\df}{f} 
\newcommand{\CovMat}[1]{{\bf C}_{#1}}

\newcommand{\MaxEig}[1]{\lambda_{\max}\!\left( #1 \right)}
\newcommand{\MinEig}[1]{\lambda_{\min}\!\left( #1 \right)}

\acrodef{mse}[MSE]{mean-squared error}
\acrodef{mmse}[MMSE]{minimal MSE}
\acrodef{lmmse}[LMMSE]{linear minimum MSE}
\acrodef{ta}[TA]{{\em time-averaged}}
\acrodef{tamse}[TA-MSE]{{\em time-averaged} MSE}
\acrodef{mtamse}[MTA-MSE]{minimum TA-MSE}
\acrodef{lmtamse}[LMTA-MSE]{linear minimal TA-MSE}
\acrodef{fir}[FIR]{finite impulse response}
\acrodef{cwf}[CWF]{cyclic Wiener filter}
\acrodef{fresh}[FRESH]{frequency shift}
\acrodef{lti}[LTI]{linear time-invariant}
\acrodef{calms}[TA-LMS]{time-averaged LMS}
\acrodef{lptv}[LPTV]{linear periodically time-varying}
\acrodef{lms}[LMS]{least mean-squares}
\acrodef{dt}[DT]{discrete-time}
\acrodef{sd}[SD]{steepest descent}
\acrodef{soi}[SOI]{signal of interest}
\acrodef{apa}[APA]{affine projection algorithm}
\acrodef{snr}[SNR]{signal-to-noise ratio}
\acrodef{msd}[MSD]{mean-square deviation}
\acrodef{ofdm}[OFDM]{orthogonal frequency division multiplexing}
\acrodef{qam}[QAM]{quadrature amplitude modulated}
\acrodef{plc}[PLC]{power line communications}
\acrodef{nb}[NB]{narrowband}
\acrodef{ctf}[CTF]{channel transfer function}
\acrodef{acgn}[ACGN]{additive cyclostationary Gaussian noise}
\acrodef{wss}[WSS]{wide-sense stationary}
\acrodef{pdf}[PDF]{probability density function}
\acrodef{rhs}[RHS]{right hand side}
\acrodef{jwss}[JWSS]{jointly wide-sense stationary}
\acrodef{wscs}[WSCS]{wide-sense cyclostationary}
\acrodef{jwscs}[JWSCS]{jointly \ac{wscs}}
\acrodef{pc}[PC]{proper-complex}
\acrodef{jpc}[JPC]{jointly proper-complex}
\acrodef{etamse}[ETA-MSE]{excess TA-MSE}

\ifsinglecol

\else

\fi 

\long\def\symbolfootnote[#1]#2{\begingroup\def\thefootnote{\fnsymbol{footnote}}\footnote[#1]{#2}\endgroup}

\definecolor{NewColor}{rgb}{0,0,0}
\definecolor{NewColor2}{rgb}{0.5,0,0.5}

 \IEEEoverridecommandlockouts
\ifsinglecol
\fi 
\title{Performance Analysis of LMS Filters with non-Gaussian Cyclostationary Signals 
}

\vspace{-0.5cm}

\author{
\IEEEauthorblockN{ Nir Shlezinger and Koby Todros\\
}

\thanks{N. Shlezinger is with the EE department, Technion - Israel Institute of Technology, Haifa, Israel (nirshlezinge@technion.ac.il).}
\thanks{K. Todros is with the ECE department, Ben-Gurion University of the Negev, Be'er-Sheva, Israel (todros@ee.bgu.ac.il).}

\vspace{-1.4cm}

}

\vspace{-0.75cm}

\begin{document}

\maketitle

\begin{abstract}
The least mean-square (LMS) filter is one of the most common adaptive linear estimation algorithms.
In many practical scenarios, e.g., digital communications systems, the signal of interest (SOI) and the input signal are jointly wide-sense cyclostationary. 
Previous works analyzing the performance of LMS filters for this important case assume {\em Gaussian} probability distributions of the considered signals. In this work, we provide a transient and steady-state performance analysis that applies to non-Gaussian cyclostationary signals. In the considered analysis, the SOI is modeled as a perturbed response of a linear periodically time-varying system to the input signal. We obtain conditions for convergence and derive analytical expressions for the non-asymptotic and steady-state mean-squared error. 
\textcolor{NewColor}{
We then show how the theoretical analysis can be effectively applied for two common non-Gaussian classes of input distributions, the class of compound Gaussian distributions, and the class of Gaussian mixture distributions.} 
The accuracy of our analysis is illustrated for system identification and signal recovery scenarios that involve {\em linear periodically time variant systems} and {\em non-Gaussian} inputs.

\end{abstract}

\vspace{-0.3cm}
\section{Introduction} 
\vspace{-0.1cm}
The \ac{lms} is a widely used algorithm for adaptive linear estimation  of a \ac{soi} based on an input signal. 
The \ac{lms} algorithm is a stochastic approximation of the iterative steepest descent based implementation of the Wiener filter, when the \ac{soi} and the input signal are \ac{jwss} \cite{Widrow:76,Haykin:03,Sayed:08}. This stochastic approximation involves a simple update equation which can be implemented in practical systems with low computational complexity \cite[Ch. 9]{Haykin:03}, \cite[Ch. 10]{Sayed:08}. 
As the \ac{lms} is designed for \ac{jwss} signals, many works have been devoted to analyze its performance in this setup, see, e.g., \cite{Widrow:76, Gardner:84, Feuer:85, Rupp:93, AlNaffouri:04}, and also \cite[Ch. 9]{Haykin:03}, \cite[Ch. 24]{Sayed:08}. 
Nonetheless, man-made signals, and specifically digitally modulated communications signals, are typically {\em \ac{wscs}} \cite[Sec. 5-7]{Gardner:06}, \cite[Ch. 1]{Gardner:94}, \cite{Heath:99}. Thus, in many practical communications systems, the considered signals are \ac{jwscs} \cite[Sec. 3.6.2]{Gardner:06} rather than \ac{jwss}. Examples include interference-limited communications \cite{Yang:15, Campbell:83} and cognitive radio \cite{Hong:09}. Another important example is \ac{nb} \ac{plc} systems, where the channel input is a digitally modulated \ac{wscs} signal \cite[Ch. 5]{Ferreira:10}, the channel transfer function is modeled as an \ac{lptv} system \cite{Corripio:06}, and the additive channel noise is  a \ac{wscs} process that is mutually independent of the channel input \cite[Ch. 2]{Ferreira:10}, \cite{Evans:12}. Hence, in this case, the channel input and the noisy channel output are \ac{jwscs}.

Despite the importance of the \ac{wscs} scenario, only a few works have studied the performance of adaptive algorithms 
in the presence of \ac{jwscs} input signal and \ac{soi}. 
In \cite{McLernon:91} it was shown that  the \ac{lms} filter coefficients are mean convergent only when the step-size approaches zero. In this case, the filter coefficients converge to the minimal \ac{ta} \ac{mse} filter. 
The performance of the \ac{lms} when applied to the adaptation of \ac{fresh} filters with \ac{wscs} inputs was studied in \cite{Reed:90, Zhang:99, Ojeda:10}. 
Specifically, \cite{Reed:90} focused on interference rejection in the presence of cyclostationary digitally modulated signals; 
The work \cite{Zhang:99} proposed a scheme for blind adaptation of \ac{fresh} filters using the \ac{lms} and the recursive least-squares algorithms; 
The work \cite{Ojeda:10} studied the effect of errors in the frequency shifts on the performance of \ac{lms}-based adaptive \ac{fresh} filters with a temporally independent input signal. 
In \cite{Bershad:14}, the \ac{lms} performance  was studied for the identification of a linear system whose coefficients obey a random walk model with a \ac{wscs} Gaussian input and an additive \ac{wss} Gaussian noise. We note that when the random walk effect is negligible, the linear system considered in \cite{Bershad:14} becomes \ac{lti}. 
The work \cite{Bershad:16} analyzed the \ac{lms} performance when applied to adaptive line enhancement/cancellation for a \ac{wscs} input consisting of a Gaussian process with periodic variance plus a sine wave with random phase. 
Excluding \cite{McLernon:91}, which studied mean convergence only, all the works mentioned above assume specific signal distributions or specific models that relate the input and \ac{soi}. 
%

In the works \cite{Shlezinger:16a, Shlezinger:16} we developed a different adaptive filter for \ac{jwscs} signals based on the \ac{ta}-\ac{mse} criterion and analyzed its performance. As the adaptive filter in \cite{Shlezinger:16a, Shlezinger:16} specializes the \ac{lms} only when the signals are \ac{jwss}, the performance study conducted in \cite{Shlezinger:16a, Shlezinger:16} cannot be used to characterize the \ac{lms} behavior when the signals are \ac{jwscs}. 
In this context, it is important to note that the empirical performance of the \ac{lms} presented in \cite[Sec. V]{Shlezinger:16}, in which some of the scenarios considered correspond to practical communications scenarios, cannot be predicted by any existing analytical \ac{lms} performance study. 
This is mainly due to the fact that in \cite[Sec. V]{Shlezinger:16}, non-Gaussian signals and periodically time-varying channels (as encountered, e.g., in practical \ac{nb}-\ac{plc} systems) were considered, that do not satisfy the specific distributional and model assumptions made in existing \ac{lms} performance analysis tools. 
The lack of a reliable characterization of the behavior of \ac{lms} filters in such practical setups further motivates the analysis of the \ac{lms} performance for general \ac{jwscs} signals.


\label{txt:MainCont}
{\bf {\slshape Main Contributions}:} 
\textcolor{NewColor}{
In this work, we provide a new performance analysis  of the \ac{lms} algorithm in the presence of \ac{jwscs} input and \ac{soi}. We obtain an expression for the transient \ac{mse}, derive conditions for convergence, including sufficient conditions on the \ac{lms} step-size, and characterize the  steady-state \ac{mse}. We explicitly show how the provided analysis can be effectively applied for two common classes of non-Gaussian input distributions: the class of compound Gaussian distributions, and the class of Gaussian mixture distributions.}
Unlike \cite{Bershad:14, Bershad:16}, 
we do not assume Gaussian distributed signals.   
We only make the basic assumptions that are commonly used  in the analysis of adaptive algorithms. In the considered analysis,  the \ac{soi} is modeled as a perturbed response of an \ac{lptv} system to the input signal. We note that this representation is not imposed and arises directly from the \ac{lmmse} filter, that represents an \ac{lptv} system when the \ac{soi} and the input signal are \ac{jwscs} \cite[Ch. 17.5.1]{Giannakis:98}. 
Unlike \cite{McLernon:91}, we also analyze mean-square convergence and provide analytic expressions for the non-asymptotic and steady-state \ac{mse}. 
Furthermore, unlike \cite{McLernon:91}, in the paper, convergence of the \ac{lms} filter is characterized accurately, without small step-size approximations. This is carried out by  introducing a generalized definition for convergence using the theory of asymptotically periodic sequences \cite{Agrawal:95}.

We verify our analysis in simulation examples involving \ac{lptv} systems and non-Gaussian signals. In particular, we demonstrate the accuracy of our analysis in scenarios for which the \ac{lms} performance was not yet characterized, including \ac{lptv} system identification and practical \ac{nb}-\ac{plc} signal recovery scenarios. Our simulation results show an excellent agreement between the theoretical and empirical performance measures. 

%

The paper is organized as follows: 
Section \ref{sec:Preliminaries} states the considered problem and presents the assumed signal model, Section \ref{sec:Performance} details the transient and steady-state performance analysis.
\textcolor{NewColor}{
Section \ref{sec:Application} presents how the provided analysis can be applied for common classes of non-Gaussian distributions, }
 and Section \ref{sec:Simulations} presents simulation examples. 
Lastly, Section \ref{sec:Conclusions} provides concluding remarks. 
Complete proofs for the results stated throughout the paper are provided in the Appendix.

\vspace{-0.2cm}
\section{Preliminaries and Problem Formulation}
\label{sec:Preliminaries}
\vspace{-0.1cm}
\subsection{Notations}
\label{subsec:Pre_Notations}
\vspace{-0.1cm}
We denote column vectors with lower-case boldface letters, e.g., ${\bf{x}}$; the $k$-th element ($k \geq 0$) of a vector ${\bf{x}}$ is denoted with $({\bf{x}})_k$. 
Matrices are denoted with upper-case boldface letter, e.g., ${\bf{X}}$; the  element at the $k$-th row and the $l$-th column of ${\bf{X}}$ is denoted by $({\bf{X}})_{k,l}$. 
${\bf{I}}_n$  denotes the $n \! \times \! n$ identity matrix 
and ${\bf{0}}_{n\! \times \! m}$ denotes the all-zero $n \! \times \! m$ matrix.
Hermitian transpose, transpose, complex conjugate, and stochastic expectation are denoted by $(\cdot)^H$, $(\cdot)^T$, $(\cdot)^*$, and $\E\{ \cdot \}$, respectively. 
The  real part of $x$ is denoted by ${\mathop{\rm Re}\nolimits} \left\{x\right\}$, 
$((n))_m$  denotes the remainder of $n$ when divided by $m$,
and $\otimes$  denotes the Kronecker product. 
The set of non-negative integers is denoted by $\mathds{N}$. 
For an $n \! \times \! n$ matrix ${\bf X}$,  $\MaxEig{\bf X}$  and $\MinEig{\bf X}$
 denote the largest and the smallest real eigenvalues of ${\bf X}$, respectively, given that such exists, 
$\rho(\bf X)$  denotes the spectral radius of ${\bf X}$,
and ${\bf x} = {\rm vec}\left({\bf X}\right)$  denotes the $n^2 \! \times \! 1$ column vector obtained by stacking the columns of ${\bf X}$,
which is recovered from  ${\bf x}$ via ${\bf X} = {\rm vec}^{-1}\left({\bf x}\right)$. 
For an $n \! \times \! 1$ vector ${\bf y}$ and an $n^2 \! \times \! 1$ vector ${\bf x}$, $\left\|{\bf y}\right\|^2 \triangleq {\bf y}^H{\bf y}$  denotes the squared Euclidean norm and $\left\|{\bf y}\right\|^2_{\bf x} \triangleq {\bf y}^H{\rm vec}^{-1}\left({\bf x}\right){\bf y}$ denotes its weighted version, when ${\rm vec}^{-1}\left({\bf x}\right) $ is positive-definite. 
Lastly, for a set of $n \! \times \! n$ matrices $\left\{{\bf X}_k\right\}$ and integers $l,m$,  $\prod\limits_{k = l}^{m}\! {\bf X}_k$ is the product $ {\bf X}_{m}{\bf X}_{m-1}\!\cdots{\bf X}_{l}$ when $m\! \geq\! l$ and ${\bf I}_n$ when $m < l$.

\vspace{-0.15cm}
\subsection{Wide-Sense Cyclostationary Stochastic Processes}
\label{subsec:Pre_Cyclostationarity}
\vspace{-0.1cm}
A  discrete-time \ac{pc} multivariate process ${\bf x}[n]$ is said to be \ac{wscs} if both its mean $\E\big\{{\bf x}[n]\}$ and autocorrelation function $\E\big\{{\bf x}[n \!+\! l]{\bf x}^H\![n]\}$ are periodic with some period, $N_0$, with respect to $n$ \cite[Sec. 3.5]{Gardner:06}.
%
A pair of \ac{jpc}   processes ${\bf x}_1[n], {\bf x}_2[n]$ are said to be  \ac{jwscs} with period $N_0$ if each process is \ac{wscs} with period $N_0$ and their cross-correlation function $\E\big\{{\bf x}_1[n\! + \!l]{\bf x}_2^H[n]\}$ is periodic with period $N_0$ with respect to. $n$ \cite[Sec. 3.6.2]{Gardner:06}. Note that when ${\bf x}_1[n], {\bf x}_2[n]$ are \ac{wscs} with different periods, say $N_1$ and $N_2$, and their cross-correlation function is periodic with period $N_{1,2}$, then they are also \ac{jwscs} with a period which equals the least common multiple of $N_1$, $N_2$ and $N_{1,2}$.

\vspace{-0.15cm}
\subsection{Problem Formulation}
\label{subsec:Pre_Problem}
\vspace{-0.1cm}
We wish to characterize the performance of the \ac{lms} filter for linear estimation of a scalar  \ac{soi} $d[n]$ based on an $M \times 1$ multivariate input signal ${\bf x}[n]$, where ${\bf x}[n]$ and $d[n]$ are zero-mean, \ac{jpc}, and \ac{jwscs} with period $N_0$. 
Let ${\bf h}[n]$ denote the $M \times 1$ random coefficients vector  of the LMS filter at time instance $n$. 
For a step-size $\mu$ and initial guess ${\bf h}[0]$, the \ac{lms} update equation is given by \cite[Eq. 9.5]{Haykin:03}
\vspace{-0.1cm}
\begin{equation}
\label{eqn:LMSRecursion1}
{\bf h}[n+1] = {\bf h}[n] + \mu\cdot{\bf x}[n]\left(d[n]-\hat{d}[n]\right)^*, \qquad n\geq 0, 
\vspace{-0.1cm}
\end{equation}
where $\hat{d}[n] \triangleq {\bf h}^H[n]{\bf x}[n]$  is the linear estimate of the \ac{soi} at time instance $n$. 

In order to characterize the performance of the \ac{lms} algorithm,  we formulate the relationship between $d[n]$ and ${\bf x}[n]$ using the \ac{lmmse} estimator. 
To that aim, let ${\bf h}_M[n]$ denote the deterministic coefficients vector of the \ac{lmmse} estimator of $d[n]$ based on ${\bf x}[n]$. Clearly, ${\bf h}_M[n]$ satisfies the Wiener-Hopf equations \cite[Ch. 3.5]{Sayed:08}  at each time instance $n$. Thus, letting $v[n]$ be the estimation error of the  \ac{lmmse} filter, the \ac{soi} $d[n]$ can be written as 
\vspace{-0.2cm}
\begin{equation}
\label{eqn:CycModel1}
d[n] = {\bf h}_M^H[n]{\bf x}[n] + v[n].
\vspace{-0.1cm}
\end{equation}
%
By the orthogonality principle \cite[Ch. 4.2]{Sayed:08}, the estimation error of the \ac{lmmse} filter is orthogonal to the input signal, i.e., $\E\{{\bf x}[n]v^*[n]\} = {\bf 0}_{M \times 1}$. 
Note that we do not assume a specific distribution on the input signal ${\bf x}[n]$ in \eqref{eqn:CycModel1}.

The fact that the \ac{soi} $d[n]$ and the input signal ${\bf x}[n]$ are zero-mean, \ac{jpc}, and \ac{jwscs}, results in the following properties of the \ac{lmmse} filter coefficients vector ${\bf h}_M[n]$ and the corresponding estimation error $v[n]$: 
\begin{enumerate}
\item Since $d[n]$ and ${\bf x}[n]$ are \ac{jwscs} with period $N_0$, it follows from \cite[Ch. 17.5.1]{Giannakis:98} that ${\bf h}_M\![n]$ defines an $M\! \times\! 1$ periodic sequence, i.e., the \ac{lmmse} filter represents an \ac{lptv} system.
%
\item As $d[n]$ and ${\bf x}[n]$ are  also \ac{jpc} and zero-mean, it follows from \eqref{eqn:CycModel1} that $v[n]$ is a  \ac{pc}, zero-mean, \ac{wscs} process, whose variance $\SigV[n] \triangleq \E\{|v[n]|^2\}$ is periodic with period $N_0$. 
\end{enumerate}

Similarly to the standard approach used for analyzing the \ac{lms} algorithm for \ac{jwss} signals, e.g., \cite{Haykin:03, Sayed:08, Widrow:76, Gardner:84, Feuer:85}, we make the following assumptions on the signals in \eqref{eqn:CycModel1}:
%
\ifsinglecol
\begin{enumerate}[label={\bf{\em AS\arabic*}}] 	
\else
\begin{enumerate}[leftmargin=0cm,itemindent=1cm,labelsep=.5cm, label={\bf{\em AS\arabic*}}] 	
\fi 
	\item  \label{itm:assm0} 
		The estimation error of the \ac{lmmse} estimator, $v\left[n_1\right]$, and the input signal, ${\bf x}\left[n_2\right]$, are {\em mutually independent} $\forall n_1, n_2$, see also \cite[Ch. 15.2]{Sayed:08}, \cite[Sec. B.2]{Gardner:84}. 
		This assumption is satisfied, e.g., when  $d[n] \!=\! {\bf f}_{N_0}^H[n]{\bf x}[n]\! +\! z[n]$ where ${\bf f}_{N_0}[n]$  is a deterministic \ac{lptv} filter  and $z[n]$ is a \ac{pc} \ac{wscs} process independent of ${\bf x}[n]$. 
		In this case the \ac{lmmse} filter is ${\bf h}_M\![n] = {\bf f}_{N_0}[n]$ and its estimation error is $v[n] = z[n]$. 
		This assumption also holds when $d[n]$ and  ${\bf x}[n]$ are jointly Gaussian and temporally uncorrelated.
	
	\item \label{itm:assm1} 
		The random coefficients vector ${\bf h}[n]$ in \eqref{eqn:LMSRecursion1} is independent of the instantaneous input  ${\bf x}[n]$, see  
		\cite[Pg. 392]{Haykin:03},
		\cite[Ch. 16.4]{Sayed:08}. 
		This is satisfied when, e.g., ${\bf x}[n]$ is a temporally independent process. 
	
	\item  \label{itm:assm2}	
		The  fourth-order moments of the input ${\bf x}[n]$ are bounded and periodic\footnote{\label{ftn:Assm2} A similar assumption was made in the analysis of \ac{lms} with non-Gaussian \ac{wss} inputs \cite[Ch. 24]{Sayed:08}, where it was assumed that the fourth-order moments are time-invariant \cite[Eq. (24.9)]{Sayed:08} and bounded \cite[Pg. 361]{Sayed:08}.} with period $N_0$.
		This is satisfied when, e.g., ${\bf x}[n]$ is a proper-complex \ac{wscs} process that obeys a compound-Gaussian distribution \cite{Ollila:12} with texture parameter that has a bounded second-order moment. 
\end{enumerate}

Assumptions \ref{itm:assm0}-\ref{itm:assm2} are utilized in the following section to obtain explicit convergence conditions and to derive closed-form expressions for the non-asymptotic and steady-state \ac{mse}. In Section \ref{sec:Simulations}, we show that the analysis carried out under these assumptions provides a reliable characterization of the \ac{lms} performance in practical communications scenarios, where \ref{itm:assm0}-\ref{itm:assm2} do not necessarily hold. 

\label{txt:alphaStable}
{
Finally, we emphasize that the following analysis, which characterizes performance in terms of \ac{mse}, requires the second-order statistical moments of the considered signals to be finite.  Consequently, while our analysis does not assume a specific distribution of the considered signals, it is not applicable for signals with infinite second-order statistical moments, such as $\alpha$-stable signals \cite{Talebi:17}. For filter design and \ac{lms} analysis in the presence of $\alpha$-stable signals, we refer the readers to the recent works \cite{Talebi:17,Talebi:18}.}


\vspace{-0.3cm}
\section{\ac{lms} Performance Analysis} 
\label{sec:Performance} 
\vspace{-0.1cm}
	
In the following we characterize the non-asymptotic time-evolution of the \ac{mse}, which is utilized to derive conditions for convergence. Under these conditions, we obtain an analytic expression for the steady-state \ac{mse}.   
 We emphasize that the following analysis is {\em substantially different} from the performance analysis of \cite[Sec. VI]{Shlezinger:16}. One fundamental difference is in the definition of convergence: In \cite[Sec. VI]{Shlezinger:16} the performance of a different adaptive algorithm {\em designed specifically} for \ac{jwscs} signals is analyzed. Since this algorithm is designed under the \ac{ta}-\ac{mse} criterion, the standard definitions of convergence and stability, that involve asymptotically {\em time-invariant} sequences, can be used. Unlike \cite{Shlezinger:16},  in the following we analyze the temporal \ac{mse} performance of the common \ac{lms} algorithm in the presence of \ac{jwscs} signals. Hence, as we show in Subsection \ref{subsec:Convergence}, this  requires a {\em new generalized definition of convergence} that, unlike \cite{Shlezinger:16}, involves asymptotically {\em time-variant} periodic sequences \cite{Agrawal:95}. 
\vspace{-0.2cm}
\subsection{Time-Evolution of the \ac{mse}}
\label{subsec:transient}
\vspace{-0.1cm}
In order to analyze the \ac{mse} performance of the \ac{lms} filter, we first define its instantaneous estimation error: 
\begin{equation}
\label{eqn:DefError}
e[n] \triangleq d[n] - {\bf h}^H[n]{\bf x}[n].
\end{equation}
The instantaneous \ac{mse} at time index $n$ is given by $\E\{|e[n]|^2\}$. 
To characterize the \ac{mse} time-evolution, we first obtain recursive relations for the first and (weighted) second-order statistical moments of the coefficients error vector, which, similarly to \cite[Eq. (24)]{Ojeda:10}, is defined as:
\vspace{-0.15cm}
\begin{equation}
\label{eqn:DefHBar}
\he\left[ n \right] \triangleq \ho - {\bf{h}}\left[ n \right],
\vspace{-0.1cm}
\end{equation}
where $\ho$ is the $M \times 1$ coefficients vector of the linear minimal \ac{ta}-\ac{mse} estimator  obtained from the time-averaged Wiener-Hopf equations \cite[Eq. (5)]{Ojeda:10}. 
To that aim, we define the $M \times M$ input covariance matrix 
\begin{equation}
\label{eqn:CxDef1}
{\bf{C}}_{\bf{x}}[n] \triangleq \E\left\{ {\bf{x}}\left[ n \right] {\bf{x}}^H\left[ n \right]  \right\},
\vspace{-0.1cm}
\end{equation}
%
and the $M \times 1$ vector 
\begin{equation}
\label{eqn:DefGn1}
{\bf g}[n] \triangleq {\bf h}_M[n] - \ho.
\vspace{-0.1cm}
\end{equation}
Note that ${\bf g}[n]$ represents the deviation  between the coefficients vector of the \ac{lptv} \ac{lmmse} filter ${\bf h}_M[n]$ and  the coefficients vector of the \ac{lti} minimal \ac{ta}-\ac{mse} estimator $\ho$.
A recursive relation for the expected coefficients error vector is given in the following lemma:
\begin{lemma}[Mean relation]
	\label{lem:MeanRelation} 
	The expected coefficients error vector \eqref{eqn:DefHBar} satisfies the following recursive relation for $n \geq 0$
	\vspace{-0.2cm}
	\begin{equation}
	\label{eqn:MeanRelation}
	\E\!\left\{ \he\left[ n+1 \right] \right\} \!=\! \left( {\bf I}_M - \mu \Cx[n]\right) \E\!\left\{ \he\left[ n \right] \right\} - \mu\! \cdot \!\Cx[n]{\bf g}[n].
		\vspace{-0.1cm}
	\end{equation}
	
	\noindent
	[A proof is given in Appendix \ref{app:ProofMean}]
\end{lemma}
%
The \ac{msd} in filter coefficients is defined as the stochastic expectation $\E\big\{\left\| \he\left[n \right] \right\|^2\big\}$, where $\he[n]$ is the coefficients error vector defined in \eqref{eqn:DefHBar}. Furthermore, we define the weighted \ac{msd} as $\E\left\{\big\| \he\left[n \right] \right\|_{\bf{q}}^2\big\}$, where ${\bf{q}}$ is some $M^2\times{1}$ vector, such that ${\bf Q} \triangleq {\rm vec}^{-1}\left\{\bf q\right\}$ is a Hermitian positive semi-definite matrix. 
To characterize the weighted \ac{msd}, we define the following three $M^2 \times M^2$ matrices: 
	\vspace{-0.2cm}
\begin{align}
\label{eqn:BmatDef}
\B[n] &\triangleq \E\left\{\! \left( {\bf{x}}\left[ n \right]{\bf{x}}^H\left[ n \right]\right)^T  \otimes {\bf{x}}\left[ n \right]{\bf{x}}^H\left[ n \right]  \right\},
\\
\label{eqn:FmatDef}
\F[n] &\triangleq \E\left\{\! \left(  {\bf{I}}_M \!-\! \mu {\bf{x}}\left[ n \right]{\bf{x}}^H\left[ n \right]\right) ^T \otimes  \left(  {\bf{I}}_M \!-\! \mu {\bf{x}}\left[ n \right]{\bf{x}}^H\left[ n \right]\right)  \right\},
\\
\label{eqn:PmatDef}
\Pmat[n] &\triangleq \E\left\{\! \left( {\bf{x}}\left[ n \right]{\bf{x}}^H\left[ n \right]\right)^T  \otimes \left(  {\bf{I}}_M \!-\! \mu {\bf{x}}\left[ n \right]{\bf{x}}^H\left[ n \right]\right)   \right\}.
	\vspace{-0.2cm}
\end{align}
The entries of the $M^2\times M^2$ matrices defined in \eqref{eqn:BmatDef}--\eqref{eqn:PmatDef} consist of simple transformations of second-order and fourth-order statistical moments of the input signal, and can thus be computed from these statistical moments. 
We note that  fourth-order statistical moments of the input signal were also used to characterize the performance of adaptive filters  in the analysis of the \ac{lms} performance with non-Gaussian \ac{wss} inputs, see, e.g., \cite[Eq. (24.21)]{Sayed:08}. Moreover, the fourth-order moments are explicitly characterized for various multivariate distributions, e.g., for the families of elliptical distributions \cite{Maruyama:03} and Gaussian mixture distributions \cite[Ch. 3]{McLachlan:04}.


A recursive relation for the weighted \ac{msd} is stated in the following lemma: 
\begin{lemma}[Variance relation]
	\label{lem:MeanSqRelation}
	The weighted \ac{msd} satisfies the following recursion for $n \geq 0$:
\ifsinglecol	
\vspace{-0.1cm}
	\begin{align}
	\E\!\left\{\! {\left\| \he\left[ {n\! +\! 1} \right] \right\|_{\bf{q }}^2} \right\} 
	&\!=\! 	\E\left\{ {\left\| \he\left[ n \right] \right\|_{\F[n]{\bf{q}}}^2} \right\} \! +\! {\mu ^2}{\left\| {\bf g}\left[ n \right] \right\|_{\B[n]{\bf{q}}}^2} 
	\notag \\
	&\quad\!-\!  2\mu \cdot\left({\bf g}^T[n]\!\otimes\! \E\!\left\{ \he^H\left[ n \right] \right\}\right)\Pmat[n]{\bf q} 	
	\! +\! {\mu ^2}\SigV[n]\cx^T[n]{\bf q},
	\vspace{-0.1cm}
	\label{eqn:VarRelation}
	\end{align}
\else
		\vspace{-0.2cm}
	\begin{align}
	&\E\!\left\{\! {\left\| \he\left[ {n\! +\! 1} \right] \right\|_{\bf{q }}^2} \right\} 
	\!=\! 	\E\left\{ {\left\| \he\left[ n \right] \right\|_{\F[n]{\bf{q}}}^2} \right\} \! +\! {\mu ^2}{\left\| {\bf g}\left[ n \right] \right\|_{\B[n]{\bf{q}}}^2} 
	\notag \\
	&\quad\!-\!  2\mu \cdot\left({\bf g}^T[n]\!\otimes\! \E\!\left\{ \he^H\left[ n \right] \right\}\right)\Pmat[n]{\bf q} 	
		\! +\! {\mu ^2}\SigV[n]\cx^T[n]{\bf q},
			\vspace{-0.1cm}
	\label{eqn:VarRelation}
	\end{align}
\fi 
	 where  $\cx[n] \triangleq  {\rm{vec}}\left( \Cx[n] \right)$.
	
	\noindent
	[A proof is given in Appendix \ref{app:Proof2}]
\end{lemma}
%
%
Notice that for a period $N_0\!=\!1$, the input signal and the \ac{soi} are \ac{jwss}. In this case, the \ac{lmmse} filter coincides with the linear minimal \ac{ta}-\ac{mse} filter $\ho$, introduced below \eqref{eqn:DefHBar}, and therefore, by Eq. \eqref{eqn:DefGn1}, the deviation vector ${\bf g}[n]$ satisfies ${\bf g}[n]\!\equiv\! {\bf 0}_{M\!\times\!1}$. 
In this case, as expected, Lemma \ref{lem:MeanRelation} coincides with the \ac{lms} mean relation for \ac{jwss} signals  \cite[Eq. (24.2)]{Sayed:08}, 
 and Lemma \ref{lem:MeanSqRelation} coincides with the \ac{lms} variance relation for \ac{jwss} signals \cite[Eq. (24.11)]{Sayed:08}. 
 Another special case of the variance relation \eqref{eqn:VarRelation} is obtained under the following scenario: 
 Consider the problem of identifying an \ac{lti} system with \ac{wscs} input ${\bf x}[n]$ whose output is corrupted by additive \ac{wscs} noise $v[n]$, uncorrelated with the input. 
Under this scenario, one can verify that the \ac{lmmse} filter is time invariant, and therefore, the deviation vector ${\bf g}[n]$ satisfies ${\bf g}[n]\!\equiv\! {\bf 0}_{M\!\times\!1}$. 
In this case, if both ${\bf x}[n]$  and  $v[n]$ are Gaussian, and  $v[n]$ is \ac{wss} (recall that \ac{wscs} processes specialize \ac{wss} processes), then, by setting ${\bf q} \!=\! {\rm vec}^{-1}\!\left({\bf I}_M \right)$,  one can verify that \eqref{eqn:VarRelation} specializes the recursive characterization of the \ac{msd} in \cite[Eq. (15)]{Bershad:14}. We note that this specialization holds only when the random walk effect in \cite{Bershad:14} is neglected, i.e., the  system considered in \cite{Bershad:14} becomes \ac{lti}\footnote{\label{ftn:Ber}While we consider the general setup of adaptive estimation of \ac{jwscs} signals, for which the \ac{lmmse} filter coefficients vary periodically in time, the work \cite{Bershad:14} studied the identification of a linear system and modeled the temporal variations in the system coefficients via a non-stationary random walk process. As a result, the variance relation in  \cite[Eq. (15)]{Bershad:14} includes an additive variance component due to these  random variations.
	Consequently, for \ac{lti} system identification, Eq. \eqref{eqn:VarRelation} specializes \cite[Eq. (15)]{Bershad:14} when this variance term is neglected, as carried out in part of the analysis conducted in \cite{Bershad:14}.}.

In the following theorem, the  recursive relations in Lemmas \ref{lem:MeanRelation}--\ref{lem:MeanSqRelation} are used to obtain an explicit characterization of the non-asymptotic \ac{mse} of \ac{lms} filters with \ac{jwscs} input and \ac{soi}:
%
\begin{theorem}[MSE time-evolution]
	\label{thm:ExMSE} 
	 The instantaneous \ac{mse} of the \ac{lms} algorithm \eqref{eqn:LMSRecursion1} satisfies 
		\vspace{-0.2cm}
	\begin{equation}
	\E\{|e[n]|^2\} 
	= \E\left\{ \left\| \he\left[ n \right] \right\|_{\cx[n]}^2 \right\} 
	+2 {\rm Re}\left\{{\bf g}^H[n]\Cx[n]\E\{\he[n]\}\right\}
	 +{\left\| {\bf g}\left[ n \right] \right\|_{\cx[n]}^2} 
	+ \SigV[n]. 
	\label{eqn:ETaMSECurve}
		\vspace{-0.1cm}
	\end{equation}
	
	\noindent
	[A proof is given in Appendix \ref{app:ProofThmMse}]
\end{theorem}
Note that the expectations   $\E\big\{\he[n]\big\}$ and $\E\big\{ \left\| \he\left[ n \right] \right\|_{\cx[n]}^2 \big\}$ in \eqref{eqn:ETaMSECurve} can be recursively computed using \eqref{eqn:MeanRelation} and \eqref{eqn:VarRelation}, respectively. 
\label{txt:Recursion}
In particular, in order to compute $\E\big\{ \left\| \he\left[ n \right] \right\|_{\cx[n]}^2 \big\}$ by evaluating \eqref{eqn:VarRelation} with ${\bf q} = \cx[n]$,  one must compute $\E\big\{ \left\| \he\left[ n-1 \right] \right\|_{\F[n-1] \cx[n]}^2 \big\}$, which in turn can also be computed using \eqref{eqn:VarRelation} by setting ${\bf q} = \F[n-1]\cx[n]$. This recursive computation is repeated until, for $n=0$, $\E\big\{ \left\| \he\left[ 0 \right] \right\|_{\bf q}^2 \big\}$ is determined by the initial guess. 
Also notice that when ${\bf g}[n]\!\equiv\! {\bf 0}_{M\!\times\!1}$, i.e., the \ac{lmmse} filter is \ac{lti}, the \ac{mse} time-evolution \eqref{eqn:ETaMSECurve} specializes the corresponding result for the \ac{lms} with \ac{jwss} signals in \cite[Pg. 363]{Sayed:08}.
\label{txt:Complexity}
Finally, it can be shown that the computational burden of computing the recurrences \eqref{eqn:MeanRelation} and \eqref{eqn:VarRelation} is of the same order as that of the corresponding recurrences used in the analysis of \ac{lms} filters with non-Gaussian \ac{wss} signals in \cite[Ch. 24]{Sayed:08}, and that for the input signals considered in the numerical study in Subsection \ref{subsec:Scenario1}, the computational burden of computing $\E\{|e[n]|^2\}$ in \eqref{eqn:ETaMSECurve}  is of the order of $O\big(n \cdot M^3 \big)$ complex multiplications.

\vspace{-0.3cm}
\subsection{Convergence Analysis and Steady-State \ac{mse}} 
\label{subsec:Convergence} 
\vspace{-0.1cm}
Here, we derive the conditions for convergence of the \ac{lms} filter with \ac{jwscs} signals, and characterize its steady-state  performance. We begin by stating the definition for asymptotically periodic sequence, which is equivalent to the one in \cite[Def. 3.1]{Agrawal:95}.
\begin{definition}[Asymptotically periodic sequence]
	\label{def:PerConv}
	 A sequence $p[n]$ is said to be {\em asymptotically periodic} with period $N_0$ if for every $k\in \{0,1,\ldots,N_0\!-\!1\} \triangleq \mathcal{N}_0$, the subsequence $p_k[n] \triangleq p[n\cdot N_0 +k]$ converges as $n \rightarrow \infty$.  
\end{definition}
\begin{remark}
	\label{rem:Comment0}
	Note that when the $N_0$ subsequences all converge to the same limit, the definition of asymptotically periodic sequences specializes the definition of convergent sequences.
\end{remark}
\begin{remark}
	\label{rem:Comment0a}
	It follows from Def. \ref{def:PerConv} that if $p[n]$ is asymptotically periodic with period $N_0$, then $p_k[n] = p[n\cdot N_0 +k]$ converges for every finite $k \in \mathds{N}$.
\end{remark}
Based on Def. \ref{def:PerConv}, we consider the following definitions for convergence:
\begin{definition}[Convergence in the mean]
	\label{def:MeanConv1}
	An adaptive filter with coefficients error vector  $\he\left[ n \right]$ is said to be {\em mean convergent} if $\E\left\{\he\left[ n \right]\right\}$ is asymptotically periodic. 
\end{definition}
\begin{definition}[Mean-square stability]
	\label{def:Stable}
	An adaptive filter with coefficients error vector  $\he\left[ n \right]$ is said to be {\em mean-square stable} if $\E\big\{\left\|\he\left[ n \right]\right\|^2\big\}$ is asymptotically periodic. 
\end{definition} 
Defs. \ref{def:MeanConv1}--\ref{def:Stable} generalize the traditional definitions for mean convergence and mean-square stability for \ac{jwss} \ac{soi} and input signal \cite[Ch. 23.2, 23.4]{Sayed:08}. These  traditional definitions require the mean coefficients error $\E\left\{\he\left[ n \right]\right\}$ to converge to ${\bf 0}_{M\times 1}$ and the \ac{msd} $\E\big\{\left\|\he\left[ n \right]\right\|^2\big\}$ to be convergent. However, as was shown in \cite[Cmt. 2]{McLernon:91}, when no specific model relating the \ac{jwscs} \ac{soi} $d[n]$ and input signal ${\bf x}[n]$ is assumed,  then $\mathop {\lim }\limits_{n \to \infty }\!\E\left\{\he\left[ n \right]\right\}\! =\! {\bf 0}_{M \times 1}$ only when the step-size $\mu \rightarrow 0$. Consequently, to be able to  {\em specify a non-infinitesimal step-size region which guarantees that the algorithm does not diverge}, we use Defs. \ref{def:MeanConv1}--\ref{def:Stable}.

To study the conditions for mean convergence, we define the following $M \times M$ matrix
\vspace{-0.2cm}
\begin{equation}
\label{eqn:RxProdDef}
\RxProd{k_1}{k_2} \triangleq \prod\limits_{l=k_1}^{N_0 -1 + k_2}\left( {\bf I}_M - \mu \Cx\left[((l))_{N_0}\right]\right).
\vspace{-0.2cm}
\end{equation}  
Using Lemma \ref{lem:MeanRelation}, we obtain the following necessary and sufficient condition for the \ac{lms} to be mean convergent:
\begin{proposition}[Necessary and sufficient condition for mean convergence]
	\label{thm:MeanCond}
	The \ac{lms} algorithm is mean convergent if and only if $\RxProd{k}{k}$, defined in \eqref{eqn:RxProdDef}, satisfies
	\vspace{-0.2cm}
	\begin{equation}
	\label{eqn:MeanCond1}
	\rho\left(\RxProd{k}{k} \right) < 1, \quad \forall k \in \mathcal{N}_0.
	\vspace{-0.1cm}
	\end{equation}
	
	\noindent
	[A proof is given in Appendix \ref{app:ProofThmMeanCond}]	
\end{proposition}
\begin{remark}
	\label{rem:Comment1}
	For a given $\mu > 0$, it is shown in Appendix \ref{app:ProofThmMeanCond} that a scenario in which $\mathop {\lim }\limits_{n \to \infty }\!\E\left\{\he\left[ n \right]\right\}\! =\! {\bf 0}_{M \times 1}$, namely, the \ac{lms} satisfies the traditional mean-convergence definition used in \cite{McLernon:91},  is when, in addition to \eqref{eqn:MeanCond1},  the deviation vector ${\bf g}[n]$ defined in \eqref{eqn:DefGn1} satisfies ${\bf g}[n] \!\equiv\! {\bf 0}_{M \times 1}$,  i.e., the \ac{lmmse} filter is \ac{lti}, as in the case of \ac{jwss} signals. However, since the \ac{lmmse} filter for \ac{jwscs} signals is \ac{lptv} \cite[Ch. 17.5.1]{Giannakis:98}, ${\bf g}[n]$ is generally non-zero. 
	Consequently, as noted in \cite[Cmt. 2]{McLernon:91}, the \ac{lms} filter with \ac{jwscs} signals generally does not satisfy traditional definition for mean convergence for any fixed step-size  $\mu > 0$. 
\end{remark}

The relation in  \eqref{eqn:MeanCond1}  states a necessary and sufficient condition for mean convergence of the \ac{lms} algorithm. 
 One scenario in which \eqref{eqn:MeanCond1} can be translated in a more explicit condition on the step-size is stated in the following corollary:
\begin{corollary}
	\label{cor:MeanCondSameEig} 
	Assume that the eigenvectors of $\Cx[k]$ are independent of $k$, i.e., $\Cx[k] = {\bf U} {\bf D}[k]{\bf U}^H$, $\forall k  \in \mathcal{N}_0$, where ${\bf U}$ is a unitary matrix (comprised of the eigenvectors)   and ${\bf D}[k]$ is a diagonal matrix. Then, the \ac{lms} algorithm is mean convergent if and only if
		\vspace{-0.2cm}	
	\begin{equation}
	\label{eqn:MeanCondSameEig}
	\mathop {\max }\limits_{m\in \{0,1,\ldots,M-1\}} \prod\limits_{l = 0}^{{N_0} - 1} {\left| {1 - \mu {{\left( {{\bf{D}}\left[ l \right]} \right)}_{m,m}}} \right|}  < 1.
	\vspace{-0.1cm}
	\end{equation}
	
	\noindent
	[A proof is given in Appendix \ref{app:MeanCondSameEig}]
\end{corollary}

Note that the scenario considered in Cor. \ref{cor:MeanCondSameEig} specializes, for example, the scenario of spatially uncorrelated input signal, i.e., when $\Cx[k]$ is a diagonal matrix, as was considered in \cite{Bershad:14}. 
In general, it is quite difficult to determine the step-size region which guarantees mean convergence from \eqref{eqn:MeanCond1}. Therefore, in the following corollary, we propose two sufficient (but not necessary) conditions on the step-size $\mu$ which guarantee mean convergence:
%
\begin{corollary}[Sufficient conditions for mean convergence]
	\label{cor:MeanSuffCond}
	The \ac{lms} algorithm  is  mean convergent if the step-size $\mu > 0$ satisfies either of the following conditions:
	\begin{subequations}
	\label{eqn:MeanSuffCond}
	\vspace{-0.2cm}
	\begin{equation}
	\label{eqn:MeanSuffConda}
	\prod\limits_{k = 0}^{{N_0} - 1} \max\left(1 - \mu \MinEig{\Cx[k]}, \mu \MaxEig{\Cx[k]} - 1 \right) < 1;	
	\vspace{-0.1cm}
	\end{equation}
	or,
	\vspace{-0.2cm}
	\begin{equation}
	\label{eqn:MeanSuffCond1}
	\mu < \frac{2}{\MaxEig{\Cx[k]}}, \quad \forall k \in \mathcal{N}_0.
		\vspace{-0.1cm}
	\end{equation}
	\end{subequations}
	
	\noindent
	[A proof is given in Appendix \ref{app:ProofCorMeanCond}]		
\end{corollary}
%
Note that \eqref{eqn:MeanSuffCond1} implies that \eqref{eqn:MeanSuffConda} is also satisfied. Therefore,  as compared to \eqref{eqn:MeanSuffConda}, and naturally to \eqref{eqn:MeanCond1}, condition \eqref{eqn:MeanSuffCond1} results in the smallest step-size region which guarantees mean convergence. However, the step-size region in \eqref{eqn:MeanSuffCond1} is stated explicitly, while \eqref{eqn:MeanSuffConda} requires finding the roots of a polynomial of order $N_0$ in $\mu$ in order to determine the step-size region.
Furthermore,  note that in the \ac{wss} case, $N_0\! =\! 1$, Eq. \eqref{eqn:MeanCond1} can be written as  $\rho\big({\bf I}_M - \mu \Cx\left[0\right]\big)\! <\! 1$, thus  \eqref{eqn:MeanSuffCond1} becomes also a necessary condition  for mean convergence, and coincides with the standard mean convergence condition for \ac{wss} signals in \cite[Ch. 24.2]{Sayed:08}.

Next, we analyze mean-square stability. To that aim, define the following $M^2 \times M^2$ matrix: 
\begin{equation}
\FProd{k_1}{k_2} \triangleq \prod\limits_{l=k_1}^{N_0 -1 + k_2}\F\left[((l))_{N_0}\right],
			\label{eqn:FProdDef}
				\vspace{-0.1cm}
\end{equation}
where $\F[n]$ is defined in \eqref{eqn:FmatDef}.
Based on the recursive relations in Lemma \ref{lem:MeanSqRelation}, we obtain the following necessary and sufficient condition for mean-square stability:
\begin{theorem}[Necessary and sufficient condition for mean-square stability]
	\label{thm:MeanSqConv}
	When the input covariance matrix $\Cx[k]$ is non-singular and its entries are bounded $\forall k \in \mathcal{N}_0$, 
	a mean convergent \ac{lms} algorithm is mean-square stable if and only if
	\vspace{-0.2cm}
	\begin{equation}
	\label{eqn:MeanSqCond1}
	\rho\left(\FProd{k}{k} \right) < 1, \quad \forall k \in \mathcal{N}_0.
	\vspace{-0.1cm}
	\end{equation}
		
	[A proof is given in Appendix \ref{app:proofMeanSqConv}]
\end{theorem}

Similarly to the necessary and sufficient condition for mean convergence in \eqref{eqn:MeanCond1}, the  necessary and sufficient condition for mean-square stability in \eqref{eqn:MeanSqCond1} requires computing the spectral radius of a product of matrices, where each matrix depends on the step-size. While it may be difficult to obtain the stability step-size region from \eqref{eqn:MeanSqCond1}, it gives rise to an explicit, yet only sufficient condition on the step-size. 
To state the sufficient condition, we define the $M^2\times M^2$ matrix
	\vspace{-0.2cm}
\begin{equation}
\A[n] \triangleq \left(\Cx^T[n]\otimes {\bf I}_M \right) + \left({\bf I}_M \otimes  \Cx[n]\right),
\label{eqn:AmatDef}
	\vspace{-0.1cm}
\end{equation}
and the $2M^2\times 2M^2$ matrix
	\vspace{-0.1cm}
\begin{equation}
{\Hmat}[n] \triangleq \frac{1}{2}\left[ {\begin{array}{*{20}{c}}
	{\A[n]}&{ - {\B[n]}}\\
	{2{{\bf{I}}_{{M^2}}}}&{{{\bf{0}}_{{M^2} \times {M^2}}}}
	\end{array}} \right],
	\vspace{-0.1cm}
\label{eqn:HmatDef}
\end{equation}
where $\B[n]$ is defined in \eqref{eqn:BmatDef}.
The sufficient condition for mean-square stability is stated in the following corollary:
\begin{corollary}[Sufficient condition for mean-square stability]
	\label{cor:MeanSqConv}
	When the input covariance matrix $\Cx[k]$ is non-singular and its entries are bounded $\forall k \in \mathcal{N}_0$, 
	a mean convergent \ac{lms} algorithm is mean-square stable if the step-size $\mu$ satisfies\footnote{If ${\Hmat}[k]$ does not have any real positive eigenvalues, then condition \eqref{eqn:MeanSqConv}  is replaced with $\mu < \frac{1}{\MaxEig{{\A^{-1}[k]}{\B[k]}}}$.} for all $k \in \mathcal{N}_0$:
		\begin{equation}
		\label{eqn:MeanSqConv} 
		\mu\! <\! \min\!\left\{\frac{1}{\MaxEig{{\A^{-1}[k]}{\B[k]}}},\frac{1}{\MaxEig{\Hmat[k]}}\right\}.
		\end{equation}

	[A proof is given in Appendix \ref{app:proofMeanSqConv2}]
\end{corollary}

Note that since  the covariance matrix  $\Cx[k]$ is assumed to be non-singular, then $\A[k]$ defined \eqref{eqn:AmatDef} is non-singular, thus \eqref{eqn:MeanSqConv}  is well-defined.  
We also note that when the signals are \ac{jwss}, i.e., $N_0=1$, then Cor. \ref{cor:MeanSqConv}  specializes the sufficient condition for stability of \ac{lms} with \ac{jwss} non-Gaussian signals in \cite[Eq. (24.24)]{Sayed:08}.

Additional, possibly stronger sufficient conditions for mean-square stability can be obtained for specific scenarios. For example, \cite{Bershad:14} characterized explicit sufficient conditions for mean-square stability of the \ac{lms} for the identification of an \ac{lti} system with random variations in the system coefficients and \ac{wss} Gaussian noise with white \ac{wscs} Gaussian inputs, by introducing additional assumptions on the rate of the variations of the statistics of the input signal. The maximal step-size values which guarantee mean-square stability in \cite[Eq. (38)-(39)]{Bershad:14} are typically larger than the corresponding values obtained using Cor. \ref{cor:MeanSqConv}, and thus constitute stronger sufficient conditions for the considered scenario.
Lastly, as stated in the following theorem, when  the conditions in  Thm. \ref{thm:MeanSqConv} 
are satisfied, and the number of iterations $n$ approaches infinity,  the transient \ac{mse} \eqref{eqn:ETaMSECurve} converges to a periodic sequence that characterizes the steady-state \ac{mse}. 
To express the steady-state \ac{mse}, we define 
\begin{equation}
\sk{k} \triangleq \left({\bf I}_{M} \! - \! \RxProd{k}{k} \right)^{\! - \!1^{\vphantom{A}}}_{\vphantom{A_A}} \sum\limits_{l=k}^{N_0-1  +k}\RxProd{l+1}{k}{\bf{C}}_{\bf{x}}\left[((l))_{N_0}\right]{\bf g}\left[((l))_{N_0}\right],
\label{eqn:skDef}
\end{equation}
and
\begin{equation}
\pk{k} \!\triangleq\! 2\Pmat^T[k]\!\left({\bf g}[k]\! \otimes\! \sk{k}^* \right)\! +\! \B^T[k]\!\left({\bf g}[k]\! \otimes\! {\bf g}^*\![k] \right) + \SigV[k]\cx[k].
\label{eqn:pkDef}
\end{equation}
 The steady-state \ac{mse} of the \ac{lms} takes the following form:
\begin{theorem}[Steady-state \ac{mse}]
	\label{thm:ssMSE}
	If the conditions stated in Thm. \ref{thm:MeanSqConv} are satisfied, then the transient \ac{mse}  \eqref{eqn:ETaMSECurve} is an asymptotically periodic sequence such that  $\xi[k] \triangleq \mathop {\lim }\limits_{n \to \infty }\E\left\{|e\left[ n\cdot N_0 + k \right]|^2\right\}$ takes the form:
		\vspace{-0.2cm}
	\begin{align}
	\xi[k]
	&= \mu^2\sum\limits_{l=k}^{N_0+k-1}\pk{((l))_{N_0}}^T\FProd{l+1}{k}\left({\bf I}_{M^2}-\FProd{k}{k}\right)^{-1}\cx[k] \notag \\
	&\quad	\!-\!2\mu{\rm Re}\left\{{\bf g}^H[k]\Cx[k]\sk{k}\right\}  +{\left\| {\bf g}\left[ k \right] \right\|_{\cx[k]}^2} 
	+ \SigV[k],
	\vspace{-0.1cm}	
	\label{eqn:ssMSE}
	\end{align}
	where $\Cx[n]$, ${\bf g}[n]$,    $\FProd{k}{l}$,  $\sk{k}$, and $\pk{k}$  are defined in \eqref{eqn:CxDef1}, \eqref{eqn:DefGn1},  \eqref{eqn:FProdDef}, \eqref{eqn:skDef}, and \eqref{eqn:pkDef},   respectively, and $\cx[n] \triangleq {\rm vec}\left( \Cx[n]\right) $

	\noindent
	[A proof is given in Appendix \ref{app:proofSteady}]	
\end{theorem}
Notice that for $N_0=1$, i.e., ${\bf x}[n]$ and $d[n]$ are \ac{jwss}, then, as noted in the discussion following Lemma \ref{lem:MeanSqRelation},  ${\bf g}[n] \equiv {\bf 0}_{M \times 1}$. In this case, the second and third summands of \eqref{eqn:ssMSE} vanish, and the first summand reduces to the excess steady-state \ac{mse} of the \ac{lms} with \ac{jwss} signals in \cite[Thm. 24.1]{Sayed:08}. Thus, Thm. \ref{thm:ssMSE} specializes the steady-state \ac{mse} characterization in \cite[Thm. 24.1]{Sayed:08} when the \ac{soi} and input are \ac{jwss}. 

\vspace{-0.3cm}
\subsection{Discussion} 
\label{subsec:Discussion} 
\vspace{-0.1cm}
First, we note that, as discussed in the previous subsection, when the signals are \ac{jwss}, the generalized performance analysis  presented in this paper coincides with the standard performance analysis of the \ac{lms} with \ac{jwss} signals in \cite[Ch. 24]{Sayed:08}. 
We further note that when the signals are not \ac{jwss}, then the vector ${\bf g}[n]$ \eqref{eqn:DefGn1}, which  represents the deviation of  the \ac{lmmse} filter from an \ac{lti} system, and essentially, the periodic dynamic of the signals (and is thus zero for the \ac{jwss} setup), has a dominant effect on the \ac{lms} behavior. 
This is observed, e.g., in the temporal statistical moments of the coefficients error characterized by the recursive relations in  \eqref{eqn:MeanRelation} and \eqref{eqn:VarRelation}, and also in the instantaneous \ac{mse} in \eqref{eqn:ETaMSECurve}. 
Consequently, the presented  performance analysis {\em quantifies the effect of these periodic dynamic} on the performance of the \ac{lms} filter, compared to the \ac{jwss} setup. 
Finally, note that the deviation vector ${\bf g}[n]$ does not appear in \cite{Bershad:14}. This is due to the fact that for the specific system model and the additive noise considered in \cite{Bershad:14}, it can be shown that the \ac{lmmse} estimator of the \ac{soi} is \ac{lti}, thus  ${\bf g}[n]$ is the zero vector.

 \color{NewColor}
\vspace{-0.3cm}
\section{Evaluation of the Theoretical Performance Under Common Non-Gaussian Distributions}
\label{sec:Application}
In order to evaluate the theoretical performance measures derived in Section \ref{sec:Performance}, one has to specify the \ac{lmmse} coefficients vector, ${\bf h}_M[n]$, and the periodic variance of the \ac{lmmse} estimation error, $\SigV[n]$, which arise from the relationship between and the input and the \ac{soi} in \eqref{eqn:CycModel1}. Furthermore, the second-order and the fourth-order statistical moments of the input signal ${\bf x}[n]$ should be known in order to compute the periodic sequences of $M^2 \times M^2$ matrices $\B[n]$, $\F[n]$, and $\Pmat[n]$, defined in \eqref{eqn:BmatDef}--\eqref{eqn:PmatDef}. Generally, the amount of parameters required to evaluate the fourth-order statistical moments, namely, $\E\left\{\! \left( {\bf{x}}\left[ n \right]\right)_{p_1}  \left( {\bf{x}}\left[ n \right]\right)^*_{p_2}  \left(  {\bf{x}}\left[ n \right] \right)_{p_3} \left( {\bf{x}}\left[ n \right]\right)^*_{p_4}   \right\}$, for each $p_1,p_2,p_3,p_4 \in \mathcal{M} = \{0,1,\ldots, M-1\}$ and $n \in \mathcal{N}_0 =  \{0,1,\ldots, N_0 -1\}$, is of the order of $O(N_0 \cdot M^4)$. Nonetheless, in the following we explicitly derive these moments for two important families of non-Gaussian multivariate distributions: $1)$ Elliptical compound Gaussian distributions \cite{Ollila:12}, and $2)$ Gaussian mixture distributions  (with a finite number of components) \cite{McLachlan:04}. Under these families, we show that the actual amount of parameters required to evaluate the fourth-order statistical moments is of the order of $O(N_0 \cdot M^2)$, a significant reduction compared to $O(N_0 \cdot M^4)$. 

\vspace{-0.3cm}
\subsection{Elliptical Compound Gaussian Inputs} 
\label{subsec:Compound} 
\vspace{-0.1cm}
 The class of compound Gaussian distributions is a common family of elliptical multivariate distributions, encountered in various applications,  including, e.g., radar systems \cite[Sec. III-E]{Ollila:12}. Some examples of compound Gaussian distributions are the multivariate $t$-distribution, the multivariate $K$-distribution, and the inverse-Gaussian-compound-Gaussian distribution \cite[Sec. IV]{Ollila:12}. A cyclostationary multivariate compound Gaussian process ${\bf x}[n]$ obeys the following stochastic representation \cite[Sec. III-E]{Ollila:12}:
\begin{equation}
\label{eqn:CompoundGauss}
{\bf x}[n] = \sqrt{\tau[n]}{\bf y}[n],
\end{equation} 
where ${\bf y}[n]$ is an $M \times 1$ multivariate proper-complex Gaussian random vector, called the {\em speckle}, with zero mean and periodic covariance $\CovMat{{\bf y}}[n]$, mutually independent of the cyclostationary scalar random variable $\tau[n]$, referred to as the {\em texture}. Consequently, the covariance matrix of ${\bf x}[n]$ is given by $\CovMat{{\bf x}}[n] = \E\{\tau[n]\} \CovMat{{\bf y}}[n]$. Furthermore,  for all $p_1,p_2,p_3,p_4 \in \mathcal{M}$, the fourth-order moments of ${\bf x}[n]$ are given by
\vspace{-0.2cm}
\begin{align}
&\E\left\{\! \left( {\bf{x}}\left[ n \right]\right)_{p_1}  \left( {\bf{x}}\left[ n \right]\right)^*_{p_2}  \left(  {\bf{x}}\left[ n \right] \right)_{p_3} \left( {\bf{x}}\left[ n \right]\right)^*_{p_4}   \right\} 
= \E\left\{\tau^2[n]\right\} \E\left\{\! \left( {\bf{y}}\left[ n \right]\right)_{p_1}  \left( {\bf{y}}\left[ n \right]\right)^*_{p_2}  \left(  {\bf{y}}\left[ n \right] \right)_{p_3} \left( {\bf{y}}\left[ n \right]\right)^*_{p_4}   \right\} \notag \\
&\quad\stackrel{(a)}{=}  \E\left\{\tau^2[n]\right\}
\bigg( 
\E\left\{\! \left( {\bf{y}}\left[ n \right]\right)_{p_1}  \left( {\bf{y}}\left[ n \right]\right)^*_{p_2} \right\}  \E\left\{\left(  {\bf{y}}\left[ n \right] \right)_{p_3} \left( {\bf{y}}^*\left[ n \right]\right)_{p_4}   \right\}  \notag \\
& \qquad \qquad \qquad \qquad
+ \E\left\{\! \left( {\bf{y}}\left[ n \right]\right)_{p_1}  \left( {\bf{y}}\left[ n \right]\right)^*_{p_4} \right\}  \E\left\{\left(  {\bf{y}}\left[ n \right] \right)_{p_3} \left( {\bf{y}}^*\left[ n \right]\right)_{p_2}   \right\}
\bigg)  \notag \\
&\quad =  \E\left\{\tau^2[n]\right\} \left(\left( \CovMat{{\bf y}}[n]\right)_{p_1, p_2} \left(  \CovMat{{\bf y}}[n]\right)_{p_3, p_4}  + \left( \CovMat{{\bf y}}[n]\right)_{p_1, p_4}  \left( \CovMat{{\bf y}}[n]\right)_{p_3, p_2} \right),
\label{eqn:FourthMoment}
\vspace{-0.2cm}
\end{align}
where $(a)$ follows from Isserlis theorem for complex Gaussian random vectors \cite[Ch. 1.4]{Koopmans:93}, combined with the proper complexity of ${\bf y}[n]$. Note that $\CovMat{{\bf x}}[n]$ and \eqref{eqn:FourthMoment} can be expressed using the periodic scalar sequences $\E\{\tau[n]\}$ and $\E\left\{\tau^2[n]\right\} $, as well as the periodic $M \times M$ matrices $\CovMat{{\bf y}}[n]$. Consequently, the overall amount of parameters required to compute these moments is of the order of $O(N_0 \cdot M^2)$.

To see how this derivation of the fourth-order moments facilitates the evaluation of the theoretical performance measures derived in Section \ref{sec:Performance}, we show how the matrices $\B[n]$,  $\F[n]$, and  $\Pmat[n]$, which are the main building blocks of the proposed analysis involving fourth-order moments, are obtained here. 
		We begin with the matrix $\B [n]$, defined in \eqref{eqn:BmatDef}, whose entries can now be written as
		\begin{align*}
		\left( {\bf{B}}\left[ n \right]\right)_{l_1 \cdot M + q_1, l_2 \cdot M + q2} 
		&= \E \left\{\left({\bf x}[n] \right)_{l_2} \left({\bf x}[n] \right)_{l_1}^*\left({\bf x}[n] \right)_{q_1}\left({\bf x}[n] \right)_{q_2}^*  \right\} \notag \\
		&=   \E\left\{\tau^2[n]\right\} \left(\left( \CovMat{{\bf y}}[n]\right)_{l_2, l_1} \left(  \CovMat{{\bf y}}[n]\right)_{q_1, q_2}  + \left( \CovMat{{\bf y}}[n]\right)_{l_2, q_2}  \left( \CovMat{{\bf y}}[n]\right)_{q_1, l_1} \right),
		\end{align*}	 
		for each $l_1, l_2, q_1, q_2 \in \mathcal{M}$. Consequently, for elliptical compound Gaussian signals, the matrix $\B [n]$ is given by
		\begin{equation*}
		\B [n] =  \E\left\{\tau^2[n]\right\} \left(  \CovMat{{\bf y}}^T[n] \otimes  \CovMat{{\bf y}}[n] + {\rm vec} \left(  \CovMat{{\bf y}}^T[n]\right) \left( {\rm vec} \left(  \CovMat{{\bf y}}[n]\right) \right)^T\right).  
		\end{equation*} 
		 Next, by using the expression for $\F[n]$ in \eqref{eqn:FmatDef} it follows that
		\begin{align*}
		{\bf{F}}\left[ n \right] 
		&=   {{\bf{I}}_{{M^2}}} - \mu \cdot \E \left\{\tau[n]\right\}  \cdot \left( \left( {\CovMat{{\bf y}}^T[n]\otimes {{\bf{I}}_M}} \right) + \left( {{\bf{I}}_M} \otimes \CovMat{{\bf y}}[n] \right)\right)  + {\mu ^2}\cdot{\bf{B}}\left[ n \right]. 
		\end{align*} 
		 Similarly, by using the expression for $\Pmat[n]$ in \eqref{eqn:PmatDef} we have that
		\begin{align*}
		{\bf{P}}\left[ n \right] 
		&= \E \left\{\tau[n]\right\}  \cdot   \left( {\CovMat{{\bf y}}^T[n]\otimes {{\bf{I}}_M}} \right)  - \mu \cdot {\bf{B}}\left[ n \right]. 
		\end{align*}
		The above derivation demonstrates how the theoretical analysis can be facilitated for the class of elliptical compound Gaussian distributions.

\vspace{-0.3cm}
\subsection{Gaussian Mixture Inputs} 
\label{subsec:GMMs} 
\vspace{-0.1cm}
\smallskip
 Here, we derive the fourth-order statistical moments of an input  whose distribution is represented by a finite mixture of Gaussian distribution \cite[Ch. 3]{McLachlan:04}. Gaussian mixture distributions are known to be able to approximate a broad range of distributions , and are used, for example, to model impulsive signals, commonly encountered in broadband \ac{plc}, see, e.g., \cite{Lin:13}.
The probability distribution of a  cyclostationary multivariate Gaussian mixture process ${\bf x}[n]$ is comprised of $n_G$ cyclostationary proper-complex Gaussian distributions with periodic covariance matrices $\{\CovMat{G_m}[n]\}_{m=1}^{n_G}$ and periodic mixture weight parameters $\{\gamma_m[n]\}_{m=1}^{n_G}$. 
The periodic covariance of ${\bf x}[n]$ is given by $\CovMat{\bf x}[n] = \sum\limits_{m=1}^{n_G}\gamma_m[n] \cdot\CovMat{G_m}[n]$.
Furthermore, by using Isserlis theorem for complex Gaussian random vectors \cite[Ch. 1.4]{Koopmans:93}, it can be shown that the fourth-order moments of ${\bf x}[n]$ satisfy
\begin{align*}
&\E\left\{\! \left( {\bf{x}}\left[ n \right]\right)_{p_1}  \left( {\bf{x}}\left[ n \right]\right)^*_{p_2}  \left(  {\bf{x}}\left[ n \right] \right)_{p_3} \left( {\bf{x}}\left[ n \right]\right)^*_{p_4}   \right\} \notag \\    	
&\qquad = \sum\limits_{m=1}^{n_G}\gamma_m[n] \cdot\left(\left( \CovMat{G_m}[n]\right)_{p_1, p_2} \left(  \CovMat{G_m}[n]\right)_{p_3, p_4}  + \left(\CovMat{G_m}[n]\right)_{p_1, p_4}  \left(\CovMat{G_m}[n]\right)_{p_3, p_2} \right),
\end{align*} 
for all $p_1,p_2,p_3,p_4 \in \mathcal{M}$, $n \in \mathcal{N}_0$. The fourth-order moments are thus explicitly obtained from the $n_G \cdot N_0$   mixture weight parameters and the $n_G \cdot$ matrices $\{\CovMat{G_m}[n]\}_{m=1}^{n_G}$. Consequently, the  overall amount of parameters required to compute the second-order and fourth-order statistical moments is of the order of $O(n_G \cdot N_0 \cdot M^2)$, which, for small $n_G$, is of the order of $O(N_0 \cdot M^2)$.	Similarly to the derivation of $\B [n]$ for elliptical compound Gaussian inputs,  the matrix $\B [n]$ for Gaussian mixture inputs can be written as
\begin{equation*}
\B [n] =  \sum\limits_{m=1}^{n_G}\gamma_m[n] \cdot \left(  \CovMat{G_m}^T[n] \otimes  \CovMat{G_m}[n] + {\rm vec} \left(  \CovMat{G_m}^T[n]\right) \left( {\rm vec} \left(  \CovMat{G_m}[n]\right) \right)^T\right).  
\end{equation*} 		
 By repeating the arguments in the previous subsection, explicit expressions for $\F [n]$ and $\Pmat [n]$ can also be obtained,  facilitating the evaluation of the theoretical performance measures derived in Section \ref{sec:Performance}.

The derivations of the statistical moments given in this section can be used to theoretically evaluate the performance for important non-Gaussian distributions encountered in various applications. Furthermore, our derivations demonstrate that for these important families of distributions, the amount of parameters required to evaluate the theoretical performance measures is of the order of $O(N_0 \cdot M^2)$, which is significantly smaller compared to the total number of elements comprising the periodic matrices   $\B[n]$, $\F[n]$, and $\Pmat[n]$, defined in \eqref{eqn:BmatDef}--\eqref{eqn:PmatDef}.

\color{black}
\vspace{-0.3cm}
\section{Numerical Examples}
\label{sec:Simulations}
\vspace{-0.1cm}
Here, we demonstrate the theoretical analysis in a simulation study, consisting of two parts: 
\begin{enumerate}
	\item First, in Subsection \ref{subsec:Scenario1}, we consider two examples which satisfy \ref{itm:assm0}-\ref{itm:assm2}, whose  purpose is to {\em verify the theoretical analysis} and to {\em numerically compare our analysis with the state-of-the-art}.
	\item Then, in Subsection \ref{subsec:Scenario2}, we consider an \ac{nb}-\ac{plc} signal recovery scenario, which {\em demonstrates the accuracy of the analysis in a practical scenario} when \ref{itm:assm0}-\ref{itm:assm2} are not necessarily satisfied. 
\end{enumerate}
All empirical performance measures were obtained  via $10000$ Monte Carlo simulations.

\vspace{-0.3cm}
\subsection{Theoretical Analysis Verification} 
\label{subsec:Scenario1} 
\vspace{-0.1cm}
In order to verify the theoretical analysis presented in Thms. \ref{thm:ExMSE}-\ref{thm:ssMSE}, we consider two examples of \ac{jwscs} input signal ${\bf x}[n]$ and \ac{soi} $d[n]$ for which \ref{itm:assm0}-\ref{itm:assm2} are satisfied. To that aim, 
let $N_x$, $N_h$, and $N_v$ be positive integers, and fix $M \! = \! 8$. 
\label{txt:Example1Start}
In the first example,
we let ${\bf t}[n]$ be an $M\times 1$ i.i.d. random process, where $\forall n \in \mathds{N}$, ${\bf t}[n]$ has a zero-mean multivariate $t$-distribution with  $\df  = 5$ degrees of freedom and scatter matrix $\CovMat{{\bf t}}$ \cite[Sec. IV]{Ollila:12}, whose entries are given by  $\left( \CovMat{{\bf t}}\right)_{k,l} =  e^{-|k-l| +j\frac{2\pi(k-l)}{M}}$, $k,l \in  \mathcal{M}$. 
The input signal ${\bf x}[n]$ is  a zero-mean \ac{wscs} process given by 	
\begin{equation*}
{\bf x}[n] \! \triangleq \! \left( 1\! + \!0.5\cos\left(\frac{2\pi n}{N_x}\right)\right)  {\bf t}[n]. 
\end{equation*} 
Since ${\bf x}[n]$ is temporally independent, it follows that \ref{itm:assm1} is satisfied.
\textcolor{NewColor}{
Note that since the multivariate $t$-distribution belongs to the class of elliptical compound Gaussian distributions, the second-order and the fourth-order moments of the input can be obtained as detailed in Subsection \ref{subsec:Compound}. Also notice that for $\df > 4$, \eqref{eqn:FourthMoment} implies that the fourth-order moments of ${\bf x}[n]$ are bounded and periodic with period $N_x$, i.e., \ref{itm:assm2} is satisfied. } 
\label{txt:Example1End}
We set the $M \times 1$ \ac{lmmse} coefficients vector ${\bf h}_M[n]$ in \eqref{eqn:CycModel1} such that
\begin{equation*}
\left( {\bf h}_M[n]\right)_k \! \triangleq \! \left(1\!+\!\frac{1 }{5 \cdot N_h}\cdot\left( \left( n\right) \right)_{N_h} \right)e^{-0.5|k|}.
\end{equation*}
Furthermore,  we set the estimation error process $v[n]$ in \eqref{eqn:CycModel1} to be a zero-mean temporally uncorrelated Gaussian process  independent of ${\bf x}[n]$, i.e., \ref{itm:assm0} is satisfied. The variance of $v[n]$ is given by
\begin{equation*}
\SigV[n] \! \triangleq \! 10^{-6}\cdot\left(1 \! + \! \frac{0.1}{\left( \left( n\right) \right)_{N_v}\!+\!1 }\right)^2.
\end{equation*} 

\smallskip
\label{txt:Example2Start}
In the second example, we let ${\bf z}[n]$ be an  $M\times 1$ i.i.d. random process, where $\forall n \in \mathds{N}$, ${\bf z}[n]$ obeys a multivariate Gaussian mixture distribution \cite[Ch. 3]{McLachlan:04} comprised of $n_G = 3$ zero-mean proper-complex Gaussians with covariance matrices $\{\CovMat{G_m}\}_{m=1}^{n_G}$ and mixture weight parameters $\{\gamma_m\}_{m=1}^{n_G}$. We set  $\left( \CovMat{G_m}\right)_{k,l} = 6\cdot e^{-m\cdot|k-l| +j\frac{2\pi(k-l)}{M}}$, $k,l \in \mathcal{M}$, and $\{\gamma_m\}_{m=1}^{n_G} = \{0.1, 0.2, 0.7\}$. The input signal ${\bf x}[n]$ is  a zero-mean \ac{wscs} process given by 
\vspace{-0.1cm}	
\begin{equation*}
{\bf x}[n] \! \triangleq \! \left( 1\! + \!\frac{\left( \left( n\right) \right)_{N_x}  }{N_x}\right)  {\bf z}[n]. 
\vspace{-0.1cm}	
\end{equation*} 
Note that \ref{itm:assm1} is satisfied as ${\bf x}[n]$ is temporally independent.
\textcolor{NewColor}{
It follows from the derivation of the fourth-order moments of Gaussian mixture distributions detailed in Subsection \ref{subsec:GMMs} that  ${\bf x}[n]$ satisfies \ref{itm:assm2}. }
\label{txt:Example2End}
In this example, the \ac{lmmse} coefficients vector ${\bf h}_M[n]$ is set to
\begin{equation*}
\left( {\bf h}_M[n]\right)_k \! \triangleq \! \left(1\!+\!\frac{1 }{100 \cdot N_h}\cdot\left( \left( n\right) \right)_{N_h} \right)\left( 1+0.1|k|\right),
\end{equation*}
and the estimation error process $v[n]$ is a zero-mean temporally uncorrelated Gaussian process  independent of ${\bf x}[n]$, thus satisfying \ref{itm:assm0}, with variance
\begin{equation*}
\SigV[n] \! \triangleq \! 10^{-6}\cdot\left(1 \! + \! {0.1} \sin\left(\frac{2 \pi n}{N_v} \right) 				\right)^2.
\end{equation*}

In both examples, the \ac{soi} $d[n]$ is related to ${\bf x}[n]$ via \eqref{eqn:CycModel1}. We conclude that $d[n]$ and ${\bf x}[n]$ are \ac{jwscs} with period $N_0$, which is the least common multiple of $N_x$, $N_h$, and $N_v$, and that these setups satisfy \ref{itm:assm0}-\ref{itm:assm2}, for any selection of  $N_x$, $N_h$, and $N_v$. 
In particular, we set $N_x=40$, $N_h=10$, and $N_v=5$.
Also note that in both examples, the input signal ${\bf x}[n]$ is non-Gaussian, the \ac{lmmse} filter ${\bf h}_M[n]$  is an \ac{lptv} system, and the estimation error $v[n]$ is a \ac{wscs} process.

Figs. \ref{fig:IID_Transient} and \ref{fig:IID_Transient2} depict the theoretical \ac{mse} computed via \eqref{eqn:ETaMSECurve}, as compared to the empirical \ac{mse},  for the first example with for step sizes $\mu = \{0.01, 0.04\}$, and for the second example with for step sizes $\mu = \{0.005, 0.01\}$, respectively. 
Figs. \ref{fig:IID_Steady} and \ref{fig:IID_Steady2} depict the theoretical steady-state \ac{ta}-\ac{mse}  for the first example and the second example, respectively,  computed by time-averaging \eqref{eqn:ssMSE} over the period $N_0$  (recall that the \ac{mse} converges to a {\em periodic} sequence), and the stability threshold. The latter is computed twice: once,
via the sufficient condition in Cor. \ref{cor:MeanSqConv}, and second, through the necessary and sufficient condition in  Thm. \ref{thm:MeanSqConv} (where the maximal step-size for which \eqref{eqn:MeanSqCond1} is satisfied is computed via grid search over $\mu$). 
Furthermore, Figs. \ref{fig:IID_Steady} and \ref{fig:IID_Steady2} also depict the empirical steady-state \ac{ta}-\ac{mse}. 
It is illustrated in  Figs. \ref{fig:IID_Transient} and \ref{fig:IID_Transient2}  that, when \ref{itm:assm0}-\ref{itm:assm2} are satisfied, the time-evolution of the \ac{mse} is accurately characterized by the theoretical analysis in Thm. \ref{thm:ExMSE}.
An excellent agreement between the empirical and theoretical performance measures is also observed in Figs. \ref{fig:IID_Steady} and \ref{fig:IID_Steady2} for both examples. One sees that, 
indeed, mean-square stability is guaranteed when the step-size satisfies \eqref{eqn:MeanSqConv}. Furthermore, note that Thm. \ref{thm:MeanSqConv} successfully identifies the empirical stability threshold.  
However, we note again that the stability threshold dictated by Thm. \ref{thm:MeanSqConv} is more difficult to find as compared to the one obtained from \eqref{eqn:MeanSqConv}. 
Lastly, we note in  Figs. \ref{fig:IID_Steady} and \ref{fig:IID_Steady2} that \eqref{eqn:ssMSE}  accurately characterizes the steady-state performance of the \ac{lms} filter after convergence is obtained. 

 \begin{figure}
 	\centering
 	\begin{minipage}{0.45\textwidth}
 		\centering
 		\scalebox{0.48}{\includegraphics{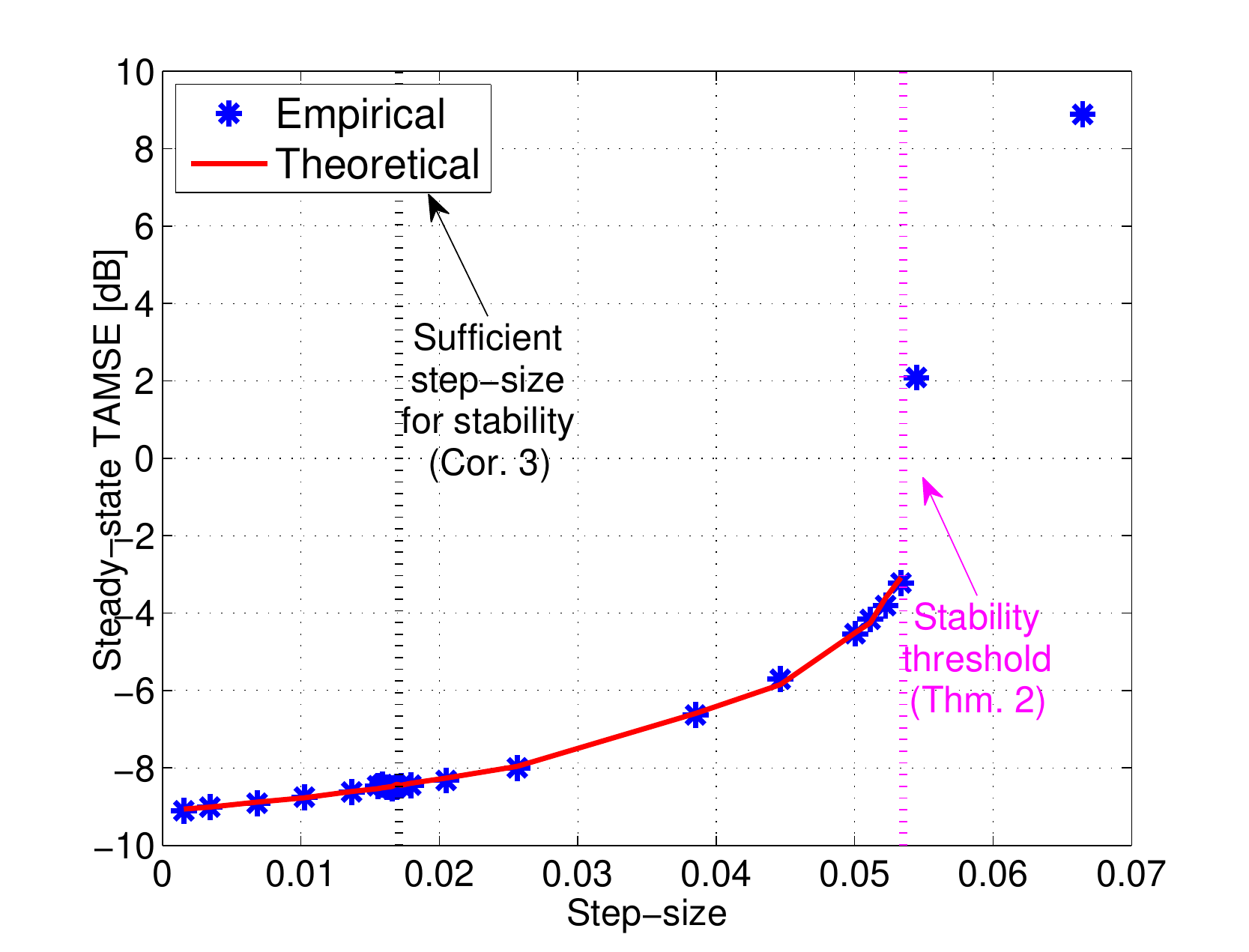}}
 		\vspace{-0.8cm}
 		\caption{The theoretical and the empirical instantaneous \ac{mse}s when \ref{itm:assm0}-\ref{itm:assm2} are satisfied, multivariate $t$-distributed input.
 		}
 		\label{fig:IID_Transient}		
 	\end{minipage}
 	$\quad$
 	\begin{minipage}{0.45\textwidth}
 		\centering
 		\scalebox{0.48}{\includegraphics{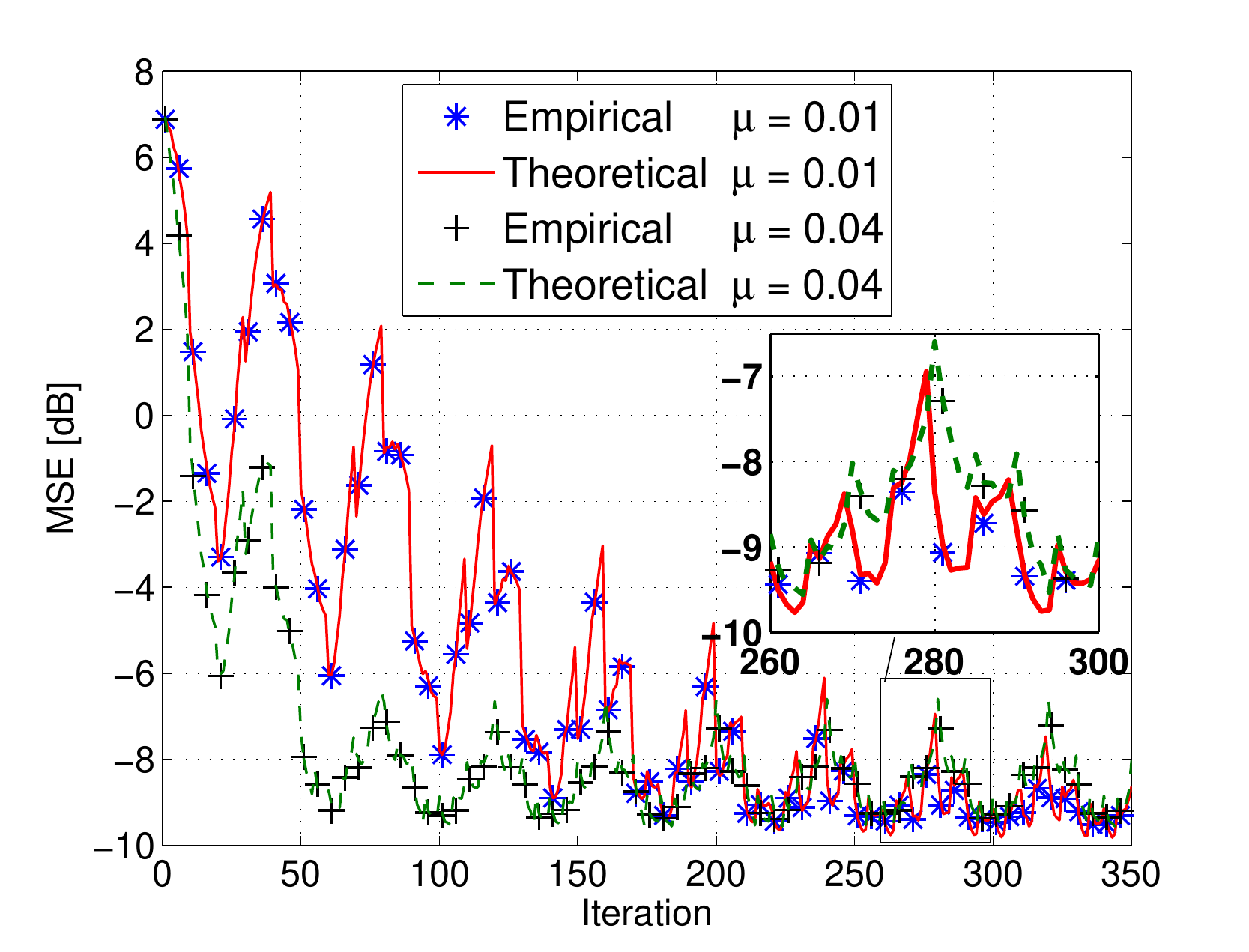}}
 		\vspace{-0.8cm}
 		\caption{The theoretical and the empirical steady-state \ac{ta}-\ac{mse}s when \ref{itm:assm0}-\ref{itm:assm2} are satisfied, multivariate $t$-distributed input.
 	}
 	\label{fig:IID_Steady}
 	\end{minipage}
 	\vspace{-0.8cm}
 \end{figure}
 \begin{figure}
 	\centering
 	\begin{minipage}{0.45\textwidth}
 		\centering
 		\scalebox{0.48}{\includegraphics{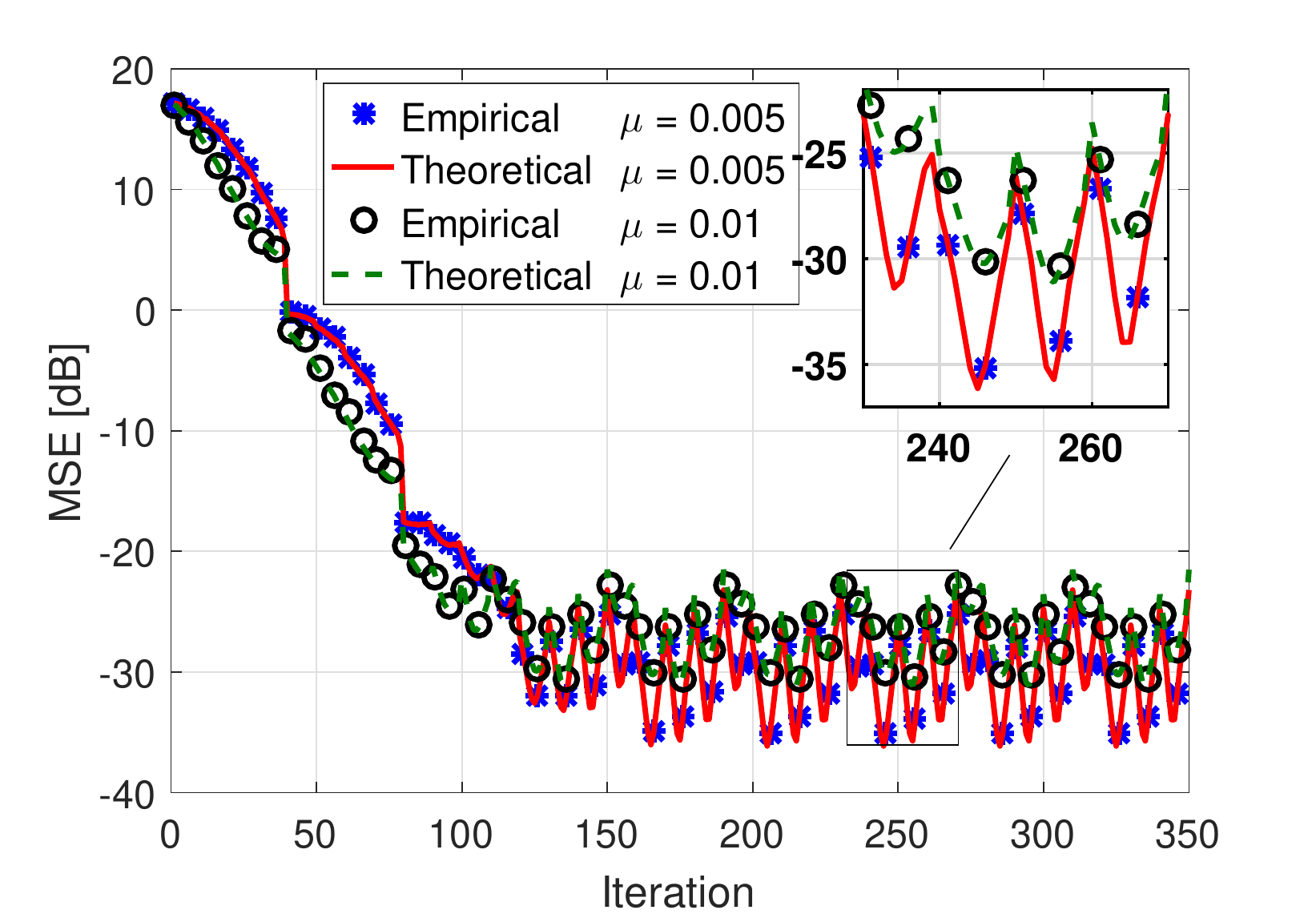}}
 		\vspace{-0.8cm}
	\caption{The theoretical and the empirical instantaneous \ac{mse}s when \ref{itm:assm0}-\ref{itm:assm2} are satisfied, multivariate Gaussian mixture input.
	}
	\label{fig:IID_Transient2}		
 	\end{minipage}
 	$\quad$
 	\begin{minipage}{0.45\textwidth}
 		\centering
 		\scalebox{0.48}{\includegraphics{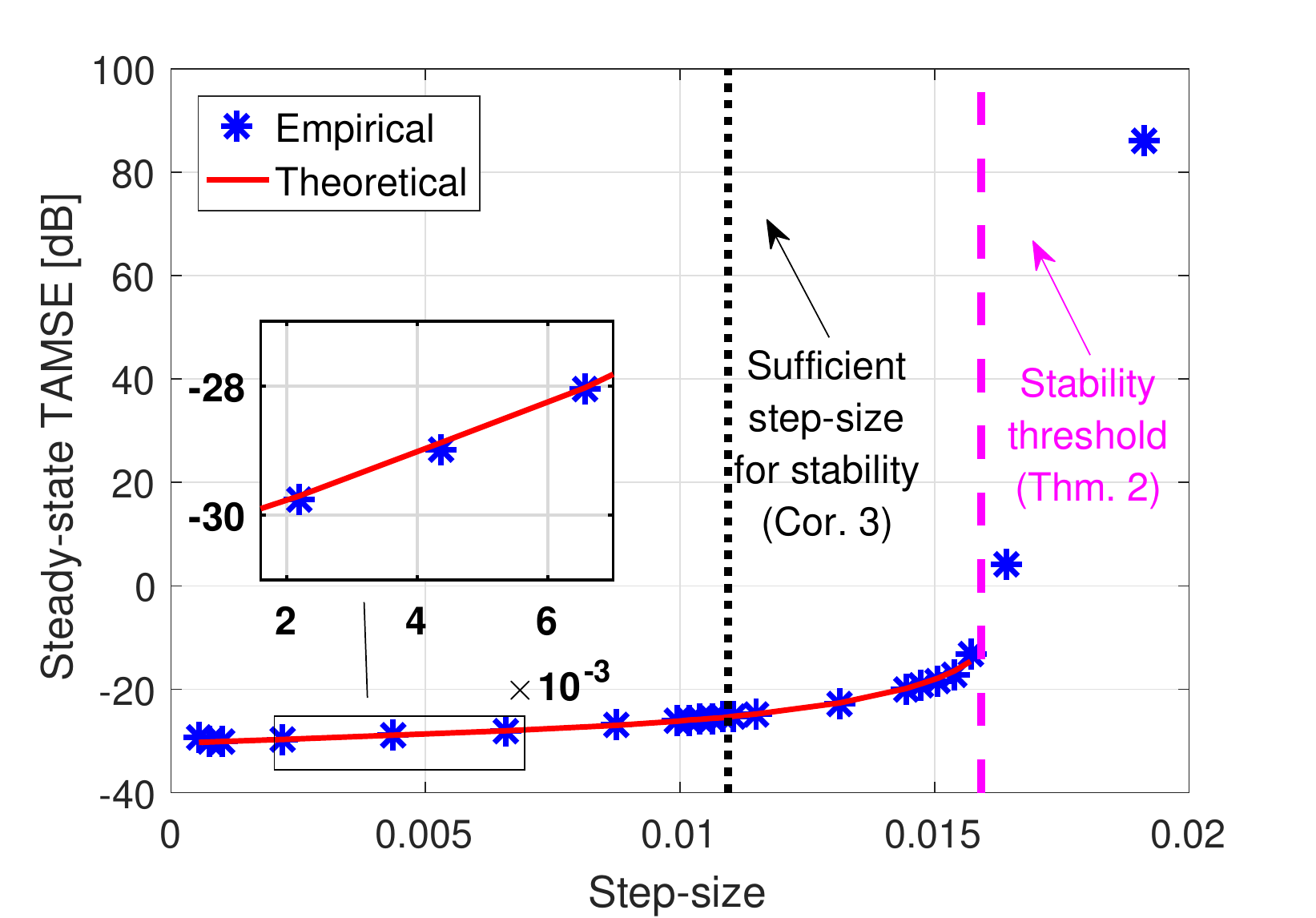}}
 		\vspace{-0.8cm}
	\caption{The theoretical and the empirical steady-state \ac{ta}-\ac{mse}s when \ref{itm:assm0}-\ref{itm:assm2} are satisfied, multivariate Gaussian mixture input.
	}
	\label{fig:IID_Steady2}
 	\end{minipage}
 	\vspace{-0.6cm}
 \end{figure}

In summary, the agreement between the empirical and theoretical performance measures observed in this  numerical study illustrate the validity of the theoretical study when  \ref{itm:assm0}-\ref{itm:assm2} are satisfied.  
 
 \vspace{-0.3cm}
 \subsection{\ac{nb}-\ac{plc} Signal Recovery} 
 \label{subsec:Scenario2} 
 \vspace{-0.1cm}
 Next, we evaluate the theoretical analysis in Thms. \ref{thm:ExMSE}-\ref{thm:ssMSE} in a practical scenario which corresponds to signal recovery in \ac{nb}-\ac{plc}  channels. 
 The \ac{soi} $d[n]$ is the input signal to an \ac{nb}-\ac{plc} channel. Here, $d[n]$ is  an \ac{ofdm} signal with $36$ subcarries, each modulated via a QPSK constellation, with $12$ cyclic prefix samples. It follows from \cite{Heath:99} that $d[n]$ is \ac{wscs} with period $N_0 \! = \! 48$. The channel output, $r[n]$, is given by $r[n] \! = \! \sum\limits_{l\! = \!0}^{\infty}g[n,l] d[n\! - \!l] \! + \! w[n]$  \cite[Sec. III]{Shlezinger:15}, where $g[n,l]$ is an \ac{lptv} filter with period $N_0$, generated as in \cite{Shlezinger:15} following the IEEE P1901.2 standard \cite{IEEE:13}. The additive channel noise $w[n]$ (note that this is not the estimation error $v[n]$ in \eqref{eqn:CycModel1}) is a \ac{wscs} Gaussian process with period\footnote{In \ac{nb}-\ac{plc} channels, the periods of the \ac{wscs} information signal, \ac{lptv} channel transfer function, and \ac{wscs} noise, are not necessarily the same \cite{Corripio:06}. However, as these periods are typically commensurate \cite{IEEE:13}, the statistical moments of the considered signals and the \ac{lptv} channel can be treated as if they have the same period, which equals the least common multiple of the different periods.} $N_0$,  generated using the model \cite{Katayama:06}, with a set of parameters taken from \cite[Tbl. 2]{Katayama:06}. 
This scenario is illustrated in Fig \ref{fig:signal_recovery1}. 
\begin{figure}
	\centering
	\includegraphics[width=0.75\textwidth]{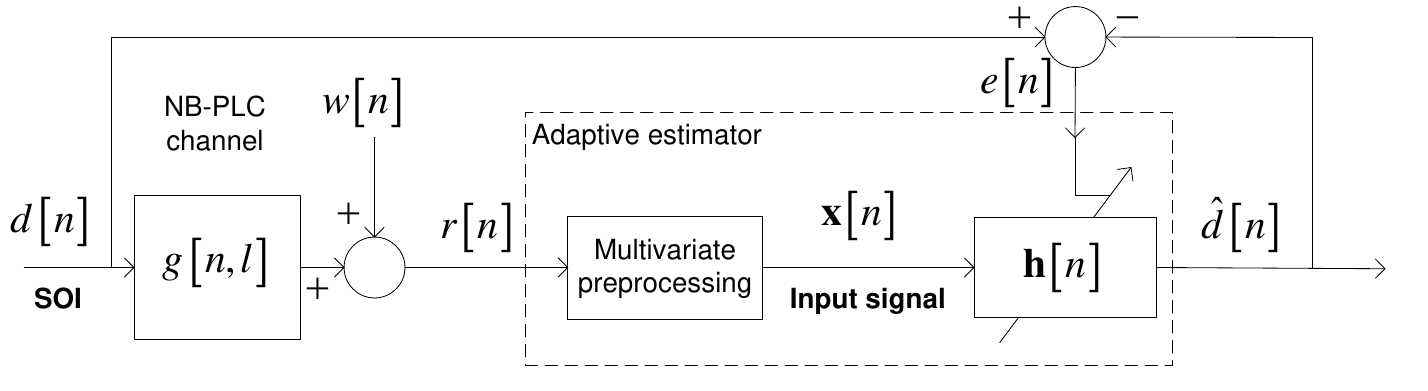}
	\vspace{-0.2cm}
	\caption{A schematic description of the \ac{nb}-\ac{plc} signal recovery scenario.}
	\vspace{-0.8cm}
	\label{fig:signal_recovery1}
\end{figure}
 The input to the \ac{lms} filter, ${\bf x}[n]$, is obtained via  multivariate preprocessing of $r[n]$ that produces  $\left({\bf x}[n] \right)_k = r[n\!-\!k]$, $k \in \{0,1,\ldots,M-1\}$, with $M\!=\!8$.  
 By representing the time-domain \ac{ofdm} signal as in \cite[Eq. (1)]{Heath:99}, it can be shown that the fourth-order statistical moments of ${\bf x}[n]$ can be analytically expressed using $g[n,l]$ and the noise parameters in  \cite[Tbl. 2]{Katayama:06}.
 The \ac{lmmse} filter ${\bf h}_M[n]$ in \eqref{eqn:CycModel1} is obtained from the orthogonality principle, i.e.,  $\E\left\{ {{\bf{x}}\left[ n \right]{d^*}\left[ n \right]} \right\} = \E\left\{ {{\bf{x}}\left[ n \right]{{\bf{x}}^H}\left[ n \right]} \right\}{\bf h}_M[n]$, and $v[n]$ is obtained as the estimation error for ${\bf h}_M[n]$. 
 We note that the entire set of assumptions \ref{itm:assm0}-\ref{itm:assm2} is not satisfied in this scenario. 
 The signal-to-noise ratio, defined as $\frac{\sum\limits_{k\!=\!0}^{N_0\!-\!1}\E\left\{ \left|r[n\!-\!k]\!-\!w[n\!-\!k]\right|^2\right\}}{ \sum\limits_{k\!=\!0}^{N_0\!-\!1}\E\left\{ \left|w[n\!-\!k]\right|^2\right\}}$,  is set to $12$ dB. 
 
 We note that the analysis in \cite{Bershad:14} is not applicable under this  scenario due to the following reasons: 
 $1)$ The work \cite{Bershad:14} is designated for system identification, while the considered \ac{nb}-\ac{plc} scenario corresponds to signal recovery; 
 $2)$ \ac{nb}-\ac{plc} channels are modeled as \ac{lptv} systems with additive \ac{wscs} noise, while in \cite{Bershad:14} a linear {\em non-periodically} time-varying system is assumed, where the temporal variations in the system coefficients obey a random walk process. Furthermore, in \cite{Bershad:14} the additive noise is a \ac{wss} process.
%

Fig. \ref{fig:PLC_Transient} depicts the empirical transient \ac{mse}  and its theoretical values, computed via \eqref{eqn:ETaMSECurve}, for step sizes $\mu = \{0.01, 0.04\}$. 
The empirical steady-state  \ac{ta}-\ac{mse} is depicted in Fig. \ref{fig:PLC_Steady}, compared to the stability threshold, computed via
 	both the sufficient condition in Cor. \ref{cor:MeanSqConv} and via the necessary and sufficient condition in  Thm. \ref{thm:MeanSqConv}, 
 and to the theoretical steady-state \ac{ta}-\ac{mse}, computed by time-averaging \eqref{eqn:ssMSE}. 
 	We note that the maximal step-size computed via the sufficient condition for stability in \eqref{eqn:MeanSqConv}, which is $\mu = 0.16$ in this example, is approximately the same as the stability threshold computed via the necessary and sufficient condition in Thm. \ref{thm:MeanSqConv}, which is $\mu = 0.163$ in this example.
The results in Fig. \ref{fig:PLC_Transient} demonstrate that even in practical scenarios where \ref{itm:assm0}-\ref{itm:assm2} are not necessarily  satisfied, there is a very good agreement between the theoretical and the empirical performance. 
Additionally, it is observed in Fig. \ref{fig:PLC_Transient} that due to the dominant periodic dynamics in the \ac{nb}-\ac{plc} scenario, the \ac{lms} exhibits significant periodic variations in the \ac{mse}.
In Fig. \ref{fig:PLC_Steady} we observe that Cor. \ref{cor:MeanSqConv} and Thm. \ref{thm:MeanSqConv}  provide a reliable prediction of the stability threshold of the \ac{lms} filter. Furthermore, one can notice that  Thm. \ref{thm:ssMSE} accurately characterizes the empirical steady-state performance, and that there is only a small gap between the theoretical and empirical measures, which arises from the fact \ref{itm:assm0}-\ref{itm:assm2} are not satisfied here. 
To conclude, these results illustrate the accuracy of the proposed analysis in practical scenarios in which assumptions \ref{itm:assm0}-\ref{itm:assm2} are not satisfied.

\begin{figure}
	\centering
	\begin{minipage}{0.45\textwidth}
		\centering
		\scalebox{0.4}{\includegraphics{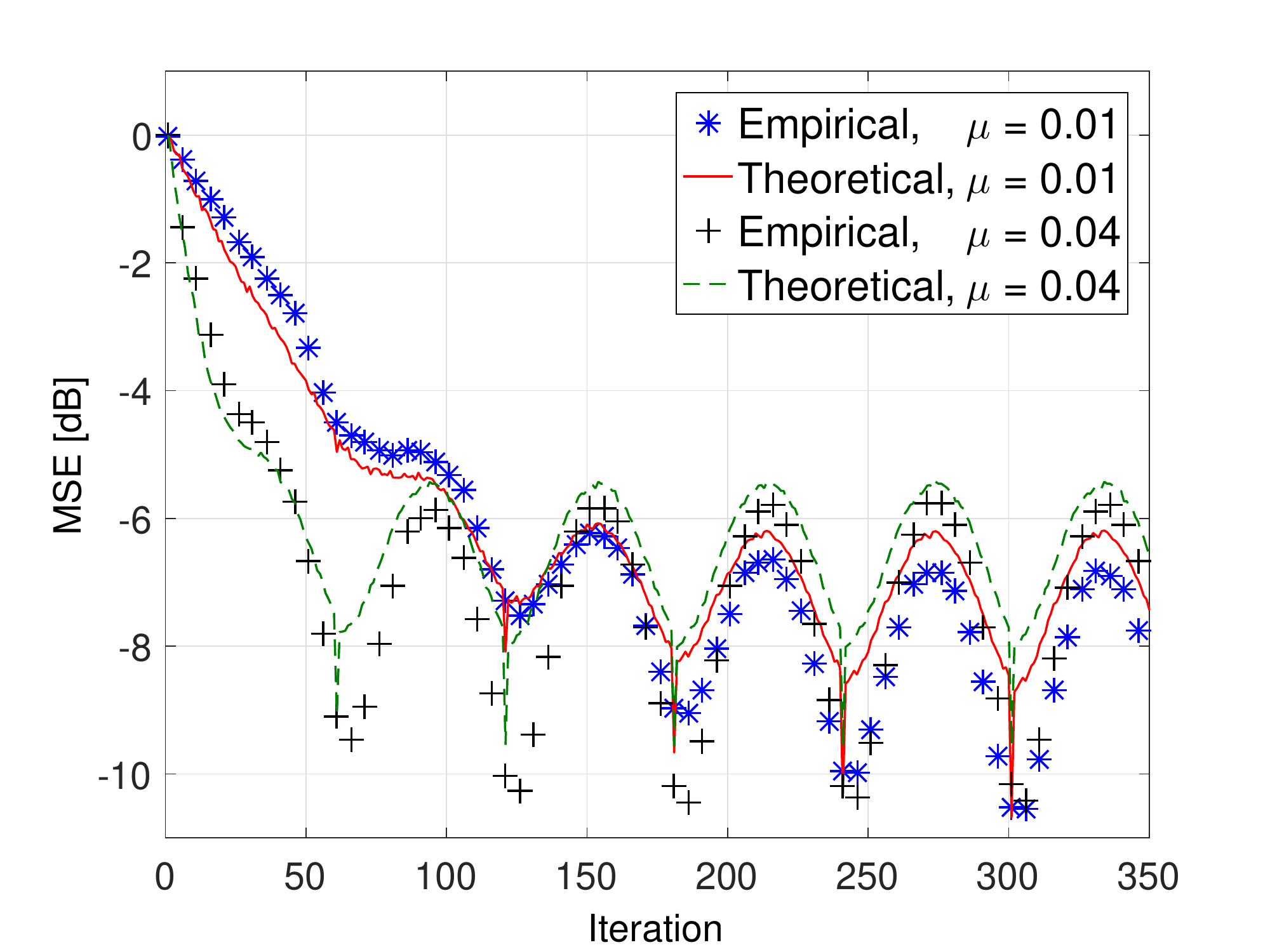}}
		\vspace{-0.8cm}
		\caption{The theoretical and the empirical instantaneous \ac{mse}s, \ac{nb}-\ac{plc} signal recovery scenario.
		}
		\label{fig:PLC_Transient}		
	\end{minipage}
	$\quad$
	\begin{minipage}{0.45\textwidth}
		\centering
		\scalebox{0.4}{\includegraphics{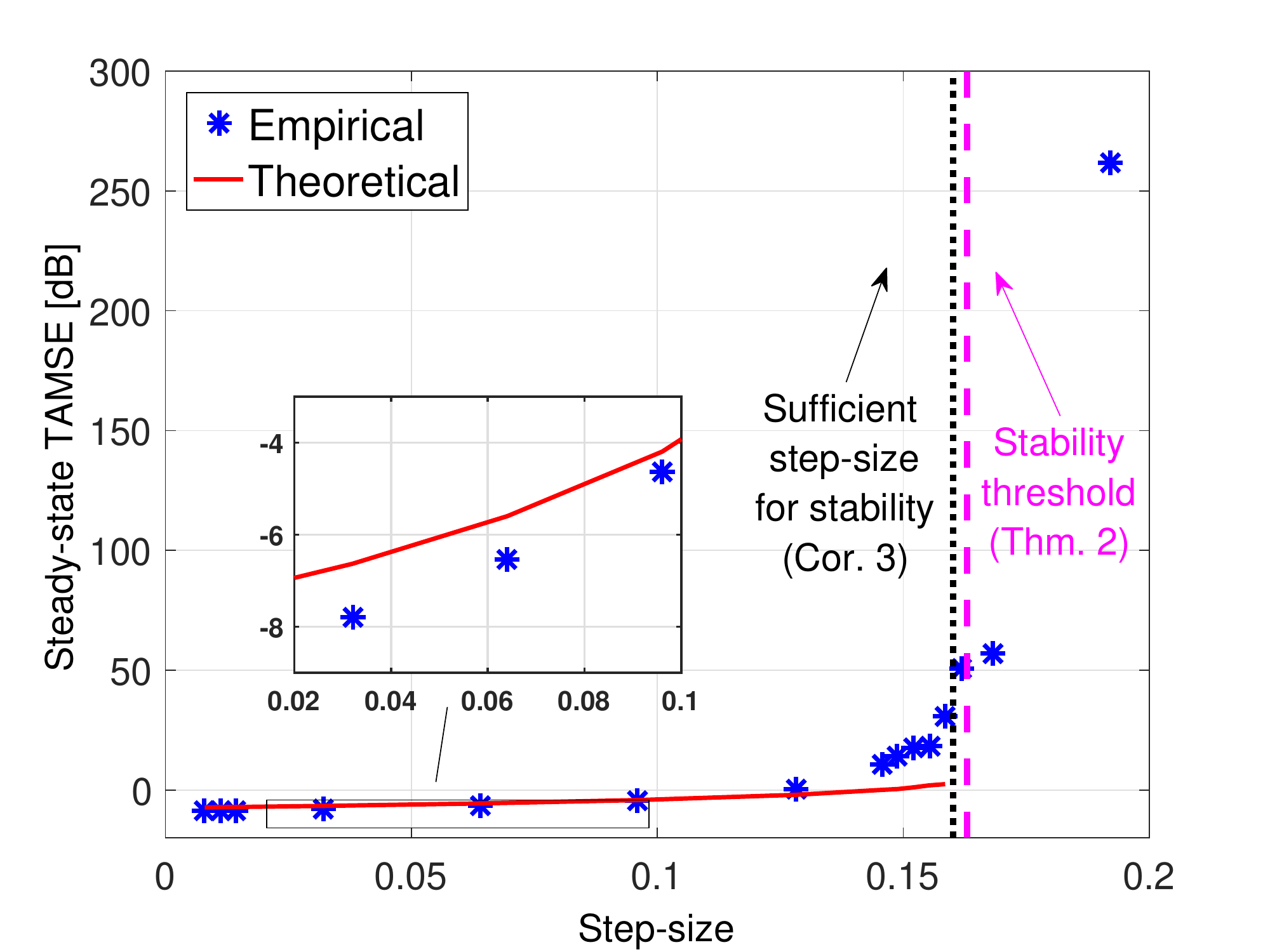}}
		\vspace{-0.8cm}
		\caption{The theoretical and the empirical steady-state \ac{ta}-\ac{mse}s, \ac{nb}-\ac{plc} signal recovery scenario.
		}
		\label{fig:PLC_Steady}
	\end{minipage}
	\vspace{-0.9cm}
\end{figure}

\vspace{-0.3cm}
\section{Conclusions}
\label{sec:Conclusions}
\vspace{-0.1cm}
This paper provides a general performance analysis of the \ac{lms} algorithm for estimation of \ac{wscs} signals. A complete characterization of the time-evolution of the \ac{mse} and the first and second-order statistical moments of the coefficients error vector was provided. Sufficient conditions for convergence and stability were obtained, and the steady-state \ac{mse} was derived. The simulation results demonstrate the accuracy of the theoretical performance measures and the stability thresholds.


\vspace{-0.3cm}
\begin{appendix}
\vspace{-0.1cm}
The following properties are repeatedly used in the sequel: 

\noindent
{\em 1)} For any matrix triplet ${\bf A}_1,{\bf A}_2, {\bf A}_3$ of compatible dimensions, it holds that (see \cite[Ch. 9.2]{Petersen:08}), 
\vspace{-0.2cm}
\begin{equation}
\label{eqn:KronVec}
{\rm vec}\left({\bf A}_1{\bf A}_2{\bf A}_3\right) = \left({\bf A}_3^T \otimes {\bf A}_1 \right){\rm vec}\left({\bf A}_2\right).
\vspace{-0.2cm}
\end{equation}
{\em 2)} For any square matrices ${\bf A}_1,{\bf A}_2$ of identical dimensions, it holds that (see \cite[Ch. 9.2]{Petersen:08}),
\vspace{-0.2cm}
\begin{equation}
\label{eqn:TrVec}
{\rm Tr}\left\{{\bf A}_1^T{\bf A}_2\right\} = {\rm vec}\left({\bf A}_1\right)^T{\rm vec}\left({\bf A}_2\right).
\vspace{-0.2cm}
\end{equation}
\noindent
{\em 3)} Since ${\bf x}[n]$ is \ac{wscs}, if follows from \ref{itm:assm2} that ${\bf{C}}_{\bf{x}}[n]$, $\B[n]$, $\F[n]$, $\Pmat[n]$, $\A[n]$, and  ${\Hmat}[n]$, defined in \eqref{eqn:CxDef1},  \eqref{eqn:BmatDef}, \eqref{eqn:FmatDef},  \eqref{eqn:PmatDef}, \eqref{eqn:AmatDef}, and \eqref{eqn:HmatDef}, respectively, are all periodic with period $N_0$.

Furthermore, we define the $M\times M$ matrix
\vspace{-0.2cm}
\begin{equation}
\Rx[n] \triangleq {\bf{I}}_M \!-\! \mu {\bf{x}}\left[ n \right]{\bf{x}}^H\left[ n \right],
\label{eqn:RxDef}
\vspace{-0.2cm}
\end{equation}
and its mean value
\vspace{-0.2cm}
\begin{equation}
\ERx[n] \triangleq \E\left\{\Rx[n]\right\} = {\bf I}_M - \mu \Cx[n].
\label{eqn:ERxDef}
\vspace{-0.2cm}
\end{equation}

\numberwithin{proposition}{subsection} 
\numberwithin{lemma}{subsection} 
\numberwithin{corollary}{subsection} 
\numberwithin{remark}{subsection} 
\numberwithin{equation}{subsection} 

\vspace{-0.2cm}
\subsection{Proof of Lemma \ref{lem:MeanRelation}}
\label{app:ProofMean}
\vspace{-0.2cm}
From \eqref{eqn:CycModel1} it follows that the instantaneous estimation error $e[n]$ \eqref{eqn:DefError} can be written as
\vspace{-0.2cm}
\begin{equation} 
\label{eqn:ErrDef2}
e[n] = \he^H\left[ n \right]{\bf{x}}\left[ n \right] + {{\bf{g}}^H}\left[ n \right]{\bf{x}}\left[ n \right] + v\left[ n \right].
\vspace{-0.2cm}
\end{equation}
From the definition of the coefficients error vector \eqref{eqn:DefHBar} it follows that $\forall n \in \mathds{N}$
\vspace{-0.2cm}
\begin{align}
\he[n\!+\!1]
&= \ho \!-\! {\bf h}[n\!+\!1] 
\stackrel{(a)}{\!=\!} \ho \!-\!{\bf h}[n] \!-\! \mu \cdot {\bf x}[n]e^*[n] \notag \\
&\stackrel{(b)}{\!=\!} \he[n] \!-\! \mu \cdot {\bf x}[n]\left(\he^H\left[ n \right]{\bf{x}}\left[ n \right] \!+\! {{\bf{g}}^H}\left[ n \right]{\bf{x}}\left[ n \right] \!+\! v\left[ n \right] \right)^*  \notag \\
&\stackrel{(c)}{\!=\!} \Rx[n]\he[n] \!-\! \mu \cdot {\bf{x}}\!\left[ n \right]\!{{\bf{x}}^H}\!\left[ n \right]\!{\bf{g}}\!\left[ n \right] \!-\! \mu \cdot {\bf{x}}\!\left[ n \right]\!{v^*}\!\left[ n \right],
\label{eqn:ProofMean1}
\vspace{-0.2cm}
\end{align}
where $(a)$ follows from \eqref{eqn:LMSRecursion1}; 
$(b)$ follows from \eqref{eqn:ErrDef2};
and $(c)$ follows from the definition of $\Rx[n]$ in \eqref{eqn:RxDef}.
Applying the stochastic expectation to \eqref{eqn:ProofMean1} yields
\vspace{-0.2cm}
\ifsinglecol
\begin{align}
\E\{\he[n\!+\!1]\} 
&= \E\left\{ \Rx[n]\he[n]\right\} \!-\! \mu \cdot \E\left\{ {\bf{x}}\left[ n \right]{{\bf{x}}^H}\right\}{\bf{g}}\left[ n \right] 
 \!-\! \mu\cdot \E\left\{{\bf{x}}\left[ n \right]{v^*}\left[ n \right]\right\} 
\notag \\
&\stackrel{(a)}{=} \ERx[n]\E\{\he[n]\} \!-\! \mu \cdot \Cx[n]{\bf g}[n],
\vspace{-0.2cm}
\end{align}
\else
\begin{align}
\E\{\he[n\!+\!1]\} 
&= \E\left\{ \Rx[n]\he[n]\right\} \!-\! \mu \cdot \E\left\{ {\bf{x}}\left[ n \right]{{\bf{x}}^H}\right\}{\bf{g}}\left[ n \right] 
\notag \\
& \quad \!-\! \mu\cdot  \E\left\{{\bf{x}}\left[ n \right]{v^*}\left[ n \right]\right\} 
\notag \\
&\stackrel{(a)}{=} \ERx[n]\E\{\he[n]\} \!-\! \mu\cdot \Cx[n]{\bf g}[n],
\vspace{-0.1cm}
\end{align}
\fi 
where $(a)$ follows since $\he[n]$ and $\Rx[n]$ are mutually independent by \ref{itm:assm1}, and since ${\bf x}[n]$ and $v[n]$ are zero-mean and mutually independent by \ref{itm:assm0}.
\qed

\vspace{-0.2cm}
\subsection{Proof of Lemma \ref{lem:MeanSqRelation}}
\label{app:Proof2}
\vspace{-0.2cm}
Set ${\bf Q} = {\rm vec}^{-1}\{{\bf q}\}$. The weighted squared Euclidean norm can be formulated as
\vspace{-0.2cm}
\ifsinglecol
\begin{align}
\left\| \he\left[ {n \!+\! 1} \right] \right\|_{\bf{q}}^2 
&= \he^H\!\left[ {n \!+\! 1} \right]{\bf{Q}}\he\left[ {n \!+\! 1} \right] \notag \\
&\stackrel{(a)}{\!=\!} {\left( \Rx\left[ n \right]\he\left[ n \right] \!-\! \mu \cdot {\bf{x}}\left[ n \right]\left( {\bf{x}}^H\!\left[ n \right]{\bf{g}}\left[ n \right] \!+\! {v^*}\left[ n \right] \right) \right)^H\!}{\bf{Q}}
\notag \\
&\quad \times 
\left( \Rx\left[ n \right]\he\left[ n \right] \!-\! \mu \cdot {\bf{x}}\left[ n \right]\left( {{\bf{x}}^H\!}\left[ n \right]{\bf{g}}\left[ n \right] \!+\! {v^*\!}\left[ n \right] \right) \right)
\notag \\
&\!=\! \he^H\!\left[ n \right]\Rx\left[ n \right]{\bf{Q}}\Rx\left[ n \right]\he\left[ n \right] 
\notag \\
&\quad 
\!+\! {\mu ^2}\cdot\left( {{{\bf{g}}^H\!}\left[ n \right]{\bf{x}}\left[ n \right] \!+\! v\left[ n \right]} \right){{\bf{x}}^H\!}\left[ n \right]{\bf{Q}}{\bf x}\left[ n \right]\left( {{{\bf{x}}^H\!}\left[ n \right]{\bf{g}}\left[ n \right] \!+\! {v^*\!}\left[ n \right]} \right)
\notag \\
&\quad \!-\! 2\mu \cdot{\rm Re} \left\{\! \he^H\!\left[ n \right]\!\Rx\!\left[ n \right]{\bf{Q}}{\bf x}\left[ n \right]\!\left( {\bf{x}}^H\!\left[ n \right]\!{\bf{g}}\!\left[ n \right] \!+\! {v^*\!}\left[ n \right] \right) \right\},
\label{eqn:ProofMeanSq1}
\vspace{-0.2cm}
\end{align}
\else 
\begin{align}
&\left\| \he\left[ {n \!+\! 1} \right] \right\|_{\bf{q}}^2 = \he^H\!\left[ {n \!+\! 1} \right]{\bf{Q}}\he\left[ {n \!+\! 1} \right] \notag \\
&\stackrel{(a)}{\!=\!} {\left( \Rx\left[ n \right]\he\left[ n \right] \!-\! \mu {\bf{x}}\left[ n \right]\left( {\bf{x}}^H\!\left[ n \right]{\bf{g}}\left[ n \right] \!+\! {v^*}\left[ n \right] \right) \right)^H\!}{\bf{Q}}
\notag \\
&\quad \times \left( \Rx\left[ n \right]\he\left[ n \right] \!-\! \mu {\bf{x}}\left[ n \right]\left( {{\bf{x}}^H\!}\left[ n \right]{\bf{g}}\left[ n \right] \!+\! {v^*\!}\left[ n \right] \right) \right)
\notag \\
&\!=\! \he^H\!\left[ n \right]\Rx\left[ n \right]{\bf{Q}}\Rx\left[ n \right]\he\left[ n \right] \notag \\
&\quad \!+\! {\mu ^2}\left( {{{\bf{g}}^H\!}\left[ n \right]{\bf{x}}\left[ n \right] \!+\! v\left[ n \right]} \right){{\bf{x}}^H\!}\left[ n \right]{\bf{Q}}{\bf x}\left[ n \right]\left( {{{\bf{x}}^H\!}\left[ n \right]{\bf{g}}\left[ n \right] \!+\! {v^*\!}\left[ n \right]} \right)
\notag \\
&\quad \!-\! 2\mu {\rm Re} \left\{\! \he^H\!\left[ n \right]\!\Rx\!\left[ n \right]{\bf{Q}}{\bf x}\left[ n \right]\!\left( {\bf{x}}^H\!\left[ n \right]\!{\bf{g}}\!\left[ n \right] \!+\! {v^*\!}\left[ n \right] \right) \right\},
\label{eqn:ProofMeanSq1}
\end{align}
\fi 
where $(a)$ follows from \eqref{eqn:ProofMean1}. 
Taking the expectation of \eqref{eqn:ProofMeanSq1}, combined with \ref{itm:assm1} yields
\ifsinglecol
\vspace{-0.2cm}
\begin{align}
\E\!\left\{ {\left\| \he\left[ {n \!+\! 1} \right] \right\|_{\bf{q}}^2} \right\} 
& \!=\! \E\!\left\{ \he^H\!\left[ n \right]\E\!\left\{ \Rx\left[ n \right]{\bf{Q}}\Rx\left[ n \right] \right\}\he\left[ n \right] \right\} \!+\!{\mu ^2}\cdot\E\!\left\{ \left| {v\left[ n \right]} \right|^2{{\bf{x}}^H\!}\left[ n \right]{\bf{Q}}{\bf x}\left[ n \right] \right\}
\notag \\
&\qquad  \!-\! 2\mu \cdot{\rm Re} \left\{  \E\!\left\{ \he^H\left[ n \right] \right\}\!\E\!\left\{ \Rx\left[ n \right]{\bf{Q}}{\bf x}\left[ n \right]{{\bf{x}}^H\!}\left[ n \right] \right\}{\bf{g}}\left[ n \right]\right\} 
\notag \\
&\qquad \!+\! {\mu ^2}\cdot{{\bf{g}}^H\!}\left[ n \right]\E\!\left\{ {{\bf{x}}\left[ n \right]{{\bf{x}}^H\!}\left[ n \right]{\bf{Q}}{\bf x}\left[ n \right]{{\bf{x}}^H\!}\left[ n \right]} \right\}{\bf{g}}\left[ n \right] .
\label{eqn:ProofMeanSq2}
\vspace{-0.2cm}
\end{align}
\else 
\begin{align}
&\E\!\left\{ {\left\| \he\left[ {n \!+\! 1} \right] \right\|_{\bf{q}}^2} \right\} \notag \\
&\qquad \!=\! \E\!\left\{ \he^H\!\left[ n \right]\E\!\left\{ \Rx\left[ n \right]{\bf{Q}}\Rx\left[ n \right] \right\}\he\left[ n \right] \right\}
\notag \\
&\qquad  \!-\! 2\mu {\rm Re} \left\{  \E\!\left\{ \he^H\left[ n \right] \right\}\!\E\!\left\{ \Rx\left[ n \right]{\bf{Q}}{\bf x}\left[ n \right]{{\bf{x}}^H\!}\left[ n \right] \right\}{\bf{g}}\left[ n \right]\right\} 
\notag \\
&\qquad \!+\! {\mu ^2}{{\bf{g}}^H\!}\left[ n \right]\E\!\left\{ {{\bf{x}}\left[ n \right]{{\bf{x}}^H\!}\left[ n \right]{\bf{Q}}{\bf x}\left[ n \right]{{\bf{x}}^H\!}\left[ n \right]} \right\}{\bf{g}}\left[ n \right]
\notag \\
&\qquad \!+\!{\mu ^2}\E\!\left\{ \left| {v\left[ n \right]} \right|^2{{\bf{x}}^H\!}\left[ n \right]{\bf{Q}}{\bf x}\left[ n \right] \right\}.
\label{eqn:ProofMeanSq2}
\end{align}
\fi 
In the following we simplify the expressions for each of the summands in \eqref{eqn:ProofMeanSq2}. 
First we note that combining \eqref{eqn:KronVec} with definitions \eqref{eqn:BmatDef}, \eqref{eqn:FmatDef}, and \eqref{eqn:PmatDef} yields
\begin{subequations}
	\label{eqn:matRel}
	\vspace{-0.2cm}
	\begin{eqnarray}
		&\E\!\left\{ \Rx\left[ n \right]{\bf{Q}}\Rx\left[ n \right] \right\} \!=\! {\rm vec}^{ \!-\! 1}\left\{ \F\left[ n \right]{\bf{q}} \right\}, 
		\label{eqn:FmatRel}\\
			&\E\!\left\{ \Rx\left[ n \right]{\bf{Q}}{\bf x}\left[ n \right]{\bf x}^H\!\left[ n \right] \right\} \!=\! {\rm vec}^{ \!-\! 1}\left\{ \B\left[ n \right]{\bf{q}} \right\}, 		
			\label{eqn:BmatRel}\\
				&\E\!\left\{ {\bf x}\left[ n \right]{\bf x}^H\!\left[ n \right]{\bf{Q}}{\bf x}\left[ n \right]{\bf x}^H\!\left[ n \right] \right\} \!=\! {\rm vec}^{ \!-\! 1}\left\{ \Pmat\left[ n \right]{\bf{q}} \right\}.
				\label{eqn:PmatRel}
	\vspace{-0.2cm}				
	\end{eqnarray}	
%
\end{subequations}
Thus, combining \eqref{eqn:KronVec} with \eqref{eqn:FmatRel} yields
\vspace{-0.2cm}
\begin{equation}
\E\!\left\{\! \he^H\!\left[ n \right]\E\!\left\{ \Rx\!\left[ n \right]\!{\bf{Q}}\Rx\!\left[ n \right] \right\}\!\he\!\left[ n \right] \right\} \!=\! \E\!\left\{\! {\left\| \he\!\left[ n \right] \right\|_{\F\left[ n \right]{\bf{q}}}^2} \right\}.
\vspace{-0.2cm}
\label{eqn:ProofSqA}
\end{equation}
Furthermore, combining \eqref{eqn:KronVec} with \eqref{eqn:PmatRel} yields 
\ifsinglecol
\vspace{-0.2cm}
\begin{equation}
\E\!\left\{ \he^H\!\left[ n \right] \right\}\E\left\{ \Rx\left[ n \right]{\bf{Q}}{\bf x}\left[ n \right]{{\bf{x}}^H\!}\left[ n \right] \right\}{\bf{g}}\left[ n \right] 
\!=\! \left( {\bf{g}}^T\left[ n \right] \otimes \E\!\left\{ \he^H\!\left[ n \right] \right\} \right)\Pmat\left[ n \right]{\bf{q}},
\vspace{-0.2cm}
\label{eqn:ProofSqB}
\end{equation}
and \eqref{eqn:KronVec} combined with \eqref{eqn:BmatRel} yields 
\vspace{-0.2cm}
\begin{equation}
{\bf{g}}^H \left[ n \right] \E \left\{  {\bf{x}}\left[ n \right]{{\bf{x}}^H }\left[ n \right] {\bf{Q}}{\bf x} \left[ n \right]{{\bf{x}}^H }\left[ n \right]  \right\}{\bf{g}}\left[ n \right]   =  \left\|  {{\bf{g}} \left[ n \right]} \right\|_{\B\left[ n \right]{\bf{q}}}^2.
\vspace{-0.2cm}
\label{eqn:ProofSqC}
\end{equation}
\else 
\begin{align}
&\E\!\left\{ \he^H\!\left[ n \right] \right\}\E\left\{ \Rx\left[ n \right]{\bf{Q}}{\bf x}\left[ n \right]{{\bf{x}}^H\!}\left[ n \right] \right\}{\bf{g}}\left[ n \right] 
\notag \\
&\qquad\qquad\qquad\!=\! \left( {\bf{g}}^T\left[ n \right] \otimes \E\!\left\{ \he^H\!\left[ n \right] \right\} \right)\Pmat\left[ n \right]{\bf{q}},
\label{eqn:ProofSqB}
\end{align}
and \eqref{eqn:KronVec} combined with \eqref{eqn:BmatRel} yields 
\begin{align}
&{\bf{g}}^H \left[ n \right] \E \left\{  {\bf{x}}\left[ n \right]{{\bf{x}}^H }\left[ n \right] {\bf{Q}}{\bf x} \left[ n \right]{{\bf{x}}^H }\left[ n \right]  \right\}{\bf{g}}\left[ n \right] \notag \\
&\qquad \qquad \qquad \qquad \qquad \qquad \qquad \quad  =  \left\|  {{\bf{g}} \left[ n \right]} \right\|_{\B\left[ n \right]{\bf{q}}}^2.
\label{eqn:ProofSqC}
\end{align}
\fi 
Lastly, from \ref{itm:assm0} it follows that 
\vspace{-0.2cm}
\begin{align}
\E\!\left\{ \left| {v\left[ n \right]} \right|^2{{\bf{x}}^H\!}\left[ n \right]{\bf{Q}}{\bf x}\left[ n \right] \right\} 
&\!=\! \SigV\left[ n \right]\E\!\left\{ {{{\bf{x}}^H\!}\left[ n \right]{\bf{Q}}{\bf x}\left[ n \right]} \right\} 
\stackrel{(a)}{\!=\!}\SigV\left[ n \right]{\bf{c}}_{\bf{x}}^T\left[ n \right]{\bf{q}},
\label{eqn:ProofSqD}
\vspace{-0.2cm}
\end{align}
where $(a)$ 
follows from \eqref{eqn:TrVec}.
Plugging \eqref{eqn:ProofSqA}, \eqref{eqn:ProofSqB}, \eqref{eqn:ProofSqC}, and \eqref{eqn:ProofSqD} into \eqref{eqn:ProofMeanSq2} yields \eqref{eqn:VarRelation}.
\qed

\vspace{-0.2cm}
\subsection{Proof of Theorem \ref{thm:ExMSE}}
\label{app:ProofThmMse}
\vspace{-0.2cm}
From \eqref{eqn:ErrDef2} it follows that
\ifsinglecol
	\vspace{-0.2cm}
\begin{align}
\E\!\left\{ {{{\left| {e\left[ n \right]} \right|}^2}} \right\} 
&= \E\!\left\{ {{{\left| \he^H\!\left[ n \right]{\bf{x}}\left[ n \right] \!+\! {{\bf{g}}^H\!}\left[ n \right]{\bf{x}}\left[ n \right] \!+\! v\left[ n \right] \right|}^2}} \right\}
\notag \\
&
\stackrel{(a)}{=}\E\!\left\{ {{{\left| {{{\left( \he\left[ n \right] \!+\! {\bf{g}}\left[ n \right] \right)}^H\!}{\bf{x}}\left[ n \right]} \right|}^2}} \right\} \!+\! \E\!\left\{ \left| {v\left[ n \right]} \right|^2 \right\}
\notag \\
&\stackrel{(b)}{=}\E\!\left\{ {{{\left| \he^H\!\left[ n \right]{\bf{x}}\left[ n \right] \right|}^2}} \right\}
 \!+\! 2{{\rm Re}} \left\{ {{{\bf{g}}^H\!}\left[ n \right]\E\!\left\{ {{\bf{x}}\left[ n \right]{{\bf{x}}^H\!}\left[ n \right]} \right\}\E\!\left\{ \he\left[ n \right] \right\}} \right\}
\notag \\
&\quad 
\!+\!{{\bf{g}}^H\!}\left[ n \right]\E\!\left\{ {{\bf{x}}\left[ n \right]{{\bf{x}}^H\!}\left[ n \right]} \right\}{\bf{g}}\left[ n \right] \!+\! \SigV\left[ n \right],
	\vspace{-0.2cm}
\label{eqn:ThmProof1}
\end{align}
\else 
\begin{align}
\E\!\left\{ {{{\left| {e\left[ n \right]} \right|}^2}} \right\} 
&= \E\!\left\{ {{{\left| \he^H\!\left[ n \right]{\bf{x}}\left[ n \right] \!+\! {{\bf{g}}^H\!}\left[ n \right]{\bf{x}}\left[ n \right] \!+\! v\left[ n \right] \right|}^2}} \right\}
\notag \\
&\stackrel{(a)}{=}\E\!\left\{ {{{\left| {{{\left( \he\left[ n \right] \!+\! {\bf{g}}\left[ n \right] \right)}^H\!}{\bf{x}}\left[ n \right]} \right|}^2}} \right\} \!+\! \E\!\left\{ \left| {v\left[ n \right]} \right|^2 \right\}
\notag \\
&\stackrel{(b)}{=}\E\!\left\{ {{{\left| \he^H\!\left[ n \right]{\bf{x}}\left[ n \right] \right|}^2}} \right\}
\notag \\
&\quad \!+\! 2{{\rm Re}} \left\{ {{{\bf{g}}^H\!}\left[ n \right]\E\!\left\{ {{\bf{x}}\left[ n \right]{{\bf{x}}^H\!}\left[ n \right]} \right\}\E\!\left\{ \he\left[ n \right] \right\}} \right\}
\notag \\
&\quad \!+\!{{\bf{g}}^H\!}\left[ n \right]\E\!\left\{ {{\bf{x}}\left[ n \right]{{\bf{x}}^H\!}\left[ n \right]} \right\}{\bf{g}}\left[ n \right] \!+\! \SigV\left[ n \right],
\label{eqn:ThmProof1}
\end{align}
\fi 
where $(a)$ follows from \ref{itm:assm0}, and $(b)$ follows from \ref{itm:assm1}.
From \ref{itm:assm1} it also follows that
	\vspace{-0.2cm}
\begin{equation}
\E\!\left\{ {{{\left| \he^H\!\left[ n \right]{\bf{x}}\left[ n \right] \right|}^2}} \right\} 
= \E\!\left\{\he^H\!\left[ n \right]\E\!\left\{ {{\bf{x}}\left[ n \right]{{\bf{x}}^H\!}\left[ n \right]} \right\}\he\left[ n \right] \right\} 
\stackrel{(a)}{=} \E\!\left\{ {\left\| \he\left[ n \right] \right\|_{{{\bf{c}}_{\bf{x}}}\left[ n \right]}^2} \right\},
	\vspace{-0.2cm}
\label{eqn:ThmProof2}
\end{equation}
where $(a)$ follows since $\cx[n] = {\rm vec}\left( \Cx[n]\right)$.
Hence, plugging \eqref{eqn:ThmProof2} into \eqref{eqn:ThmProof1} yields \eqref{eqn:ETaMSECurve}.
\qed

\vspace{-0.2cm}
\subsection{Proof of Proposition \ref{thm:MeanCond}}
\label{app:ProofThmMeanCond}
\vspace{-0.2cm}
First, as in \cite[Appendix F]{Shlezinger:16}, we define the decimated components decomposition \cite[Sec. 17.2]{Giannakis:98} of the coefficients error vector as follows: 
	\vspace{-0.2cm}
\begin{equation}
\label{eqn:hkDef}
\he_k[n] \triangleq \he[n\cdot N_0 \! + \! k], \qquad k \in \mathcal{N}_0, n \in \mathds{N}. 
	\vspace{-0.2cm}
\end{equation}
By defining 
	\vspace{-0.2cm}
	\begin{equation}
	\label{eqn:bkDef}
	\bk{k} \triangleq \sum\limits_{l=k}^{N_0-1+k}\RxProd{l+1}{k}{\bf{C}}_{\bf{x}}\left[((l))_{N_0}\right]{\bf g}\left[((l))_{N_0}\right],
		\vspace{-0.2cm}
	\end{equation}
it follows that $\forall n > N_0$, $\E\{\he_k[n\! + \!1]\} 
\! = \! \E\{\he[n\cdot N_0 \! + \! N_0 \! + \!k]\}$, and thus
\ifsinglecol
	\vspace{-0.2cm}
\begin{align}
\E\{\he_k[n\! + \!1]\} 
&\stackrel{(a)}{\! = \!}\ERx\left[((k\! - \!1))_{N_0}\right] \E\{\he[n\cdot N_0 \! + \! N_0 \! + \!k\! - \!1]\}  \! - \! \mu \!\cdot\!\Cx\left[((k\! - \!1))_{N_0}\right] {\bf g}\left[((k\! - \!1))_{N_0}\right] \notag \\
&\stackrel{(b)}{\! = \!} \RxProd{k}{k}\E\{\he_k[n]\} \! - \! \mu \!\cdot\! \bk{k},
	\vspace{-0.2cm}
\label{eqn:MeanHkRec}
\end{align}
\else 
\begin{align}
\E\{\he_k[n\! + \!1]\} 
&\! = \! \E\{\he[n\cdot N_0 \! + \! N_0 \! + \!k]\} \notag \\
&\stackrel{(a)}{\! = \!}\ERx\left[((k\! - \!1))_{N_0}\right] \E\{\he[n\cdot N_0 \! + \! N_0 \! + \!k\! - \!1]\} \notag \\
&\qquad \! - \! \mu \!\cdot\!\Cx\left[((k\! - \!1))_{N_0}\right] {\bf g}\left[((k\! - \!1))_{N_0}\right] \notag \\
&\stackrel{(b)}{\! = \!} \RxProd{k}{k}\E\{\he_k[n]\} \! - \! \mu \!\cdot\! \bk{k},
\label{eqn:MeanHkRec}
\end{align}
\fi 
where $(a)$ follows from \eqref{eqn:MeanRelation} and since $\ERx[n]$, $\Cx[n]$, and ${\bf g}[n]$, defined in \eqref{eqn:ERxDef}, \eqref{eqn:CxDef1}, and \eqref{eqn:DefGn1}, are all periodic with period $N_0$; 
$(b)$ follows from repeating the recursion $N_0$ times and plugging the expressions for $\RxProd{k}{k}$ and $\bk{k}$ defined in \eqref{eqn:RxProdDef} and \eqref{eqn:bkDef}. 
Repeating \eqref{eqn:MeanHkRec} $n$ times yields 
	\vspace{-0.2cm}
\begin{equation}
\label{eqn:MeanHk}
\!\!\E\{\he_k[n\! + \!1]\} \! = \!\left(\RxProd{k}{k} \right)^{n\! + \!1}\!\E\{\he_k[0]\} \! - \! \mu \sum\limits_{m\! = \!0}^{n}\left(\RxProd{k}{k} \right)^m\! \bk{k}.
	\vspace{-0.2cm}
\end{equation}
Therefore, $\E\big\{\he_k[n]\big\}$ converges regardless of the initial value $\E\big\{\he_k[0]\big\}$ if and only if 
$\mathop {\lim }\limits_{n \to \infty }\! \left(\RxProd{k}{k} \right)^{n} \! = \! {\bf 0}_{M \times M}$, 
%
which is satisfied if and only if $\rho\left(\RxProd{k}{k} \right) < 1$ \cite[Ch. 7.10]{Meyer:00}. This proves condition \eqref{eqn:MeanCond1}. 
Additionally, when \eqref{eqn:MeanCond1} is satisfied, then $\mathop {\lim }\limits_{n \to \infty } \sum\limits_{m\! = \!0}^{n}\left(\RxProd{k}{k} \right)^m \! = \! \left({\bf I}_{M} \! - \! \RxProd{k}{k} \right)^{\! - \!1}$  \cite[Ch. 7.10]{Meyer:00}, thus 
	\vspace{-0.2cm}
\begin{equation}
\label{eqn:MeanSSHka}
\mathop {\lim }\limits_{n \to \infty }\E\big\{\he_k[n]\big\} 
\! = \!- \mu \cdot\left({\bf I}_{M} \! - \! \RxProd{k}{k} \right)^{\! - \!1}\bk{k}.
	\vspace{-0.2cm}
\end{equation}
It follows from \eqref{eqn:MeanSSHka} that $\mathop {\lim }\limits_{n \to \infty }\E\big\{\he_k[n]\big\} \! = \! {\bf 0}_{M \times 1}$ if and only if $\bk{k}\! = \! {\bf 0}_{M \times 1}$. Noting that ${\bf g}[n]\! = \! {\bf 0}_{M \times 1}$ for all $n \in \mathcal{N}_0$ yields $\bk{k}\! = \! {\bf 0}_{M \times 1}$  proves the statement in Cmt. \ref{rem:Comment1}.
\qed

\vspace{-0.2cm}
\subsection{Proof of Corollary \ref{cor:MeanCondSameEig}}
\label{app:MeanCondSameEig}
\vspace{-0.2cm}
The corollary follows since, for $\Cx[k] = {\bf U} {\bf D}[k]{\bf U}^H$,   the matrix product $\RxProd{k}{k}$ defined in \eqref{eqn:RxProdDef} is given by $\RxProd{k}{k} = {\bf{U}}\left( {\prod\limits_{l = 0}^{{N_0} - 1} {\left( {{{\bf{I}}_M} - \mu {\bf{D}}\left[ l \right]} \right)} } \right){{\bf{U}}^H}$ for all $ k  \in \mathcal{N}_0$, and thus the spectral radius $\rho\left(\RxProd{k}{k} \right)$ in \eqref{eqn:MeanCond1} coincides with the left hand side of \eqref{eqn:MeanCondSameEig}. 
\qed

\vspace{-0.2cm}
\subsection{Proof of Corollary \ref{cor:MeanSuffCond}}
\label{app:ProofCorMeanCond}
\vspace{-0.2cm}
It follows from \eqref{eqn:RxProdDef} that $\forall k \in \mathcal{N}_0$, 
$\rho\left(\RxProd{k}{k} \right) 
 =  \rho\left(\prod\limits_{l\! = \!k}^{N_0 \! - \!1 \! + \! k}\ERx\left[((l))_{N_0}\right] \right) 
\stackrel{(a)}{\leq}
\prod\limits_{l\! = \!0}^{N_0 \! - \!1 }\rho\left(\ERx\left[l\right]\right)$,
where $(a)$ follows from \cite[Ch. 5.2]{Meyer:00}.
 Since $\ERx\left[l\right] \! = \! {\bf I}_M \! - \! \mu\! \cdot \! \Cx[l]$, where $\Cx[l]$ is Hermitian positive semi-definite, it follows that $\rho\big(\ERx\left[l\right]\big) = \max\big(1 - \mu \MinEig{\Cx[k]}, \mu \MaxEig{\Cx[k]} - 1 \big)$, thus condition \eqref{eqn:MeanSuffConda} guarantees that \eqref{eqn:MeanCond1} is satisfied. 
 
 It thus follows that condition \eqref{eqn:MeanCond1} is satisfied if (but not only if) $\rho\big(\ERx\left[l\right]\big) < 1$, $\forall l \in \mathcal{N}_0$. 
For $\mu > 0$, $\rho\big(\ERx\left[l\right]\big) < 1$ if and only if $\mu \MaxEig{\Cx[k]} - 1< 1$, proving condition \eqref{eqn:MeanSuffCond1}. 
\qed

\vspace{-0.2cm}
\subsection{Proof of Theorem \ref{thm:MeanSqConv}}
\label{app:proofMeanSqConv}
\vspace{-0.2cm}
From the proof of Prop. \ref{thm:MeanCond} it follows that if the \ac{lms} filter is mean-convergent, then $\E\big\{\he_k[n]\big\}$ converges to $- \mu \cdot\left({\bf I}_{M} \! - \! \RxProd{k}{k} \right)^{\! - \!1}\bk{k} = - \! \mu \cdot \sk{k}$, where $\he_k[n]$ is defined in \eqref{eqn:hkDef} and  $\sk{k}$ is defined in \eqref{eqn:skDef}. 
For $k \in \mathcal{N}_0$, $n \in \mathds{N}$, define 
$\tilde{\bf b}[nN_0 + k ]\triangleq \E\big\{\he_k[n]\big\} + \! \mu \cdot \sk{k}$, and
$\tilde{\bf p}[n] \triangleq \Pmat^T[n]\left({\bf g}[n]\otimes \tilde{\bf b}^*[n] \right)$.
As the entries of $\Cx[k]$ are bounded $\forall k \in \mathcal{N}_0$, it follows from \eqref{eqn:MeanHk} that if \eqref{eqn:MeanCond1} is satisfied, i.e.,  the \ac{lms} filter is mean-convergent, then the entries of $\tilde{\bf b}[n]$ are bounded $\forall n \!\in\! \mathds{N}$ and  $\mathop {\lim }\limits_{n \to \infty }\!\tilde{\bf b}[n]\! = \! {\bf 0}_{M \times 1}$. Combining this with \ref{itm:assm2} implies that the entries of  $\tilde{\bf p}[n]$ are also bounded $\forall n \!\in\! \mathds{N}$ and that  $\mathop {\lim }\limits_{n \to \infty }\!\tilde{\bf p}[n]\! = \! {\bf 0}_{M^2 \times 1}$. 

From Lemma \ref {lem:MeanSqRelation} it follows that for any $M^2 \times 1$ vector ${\bf q}$, and $\forall n > N_0$, we have that 
$\E\!\big\{\! {\left\| \he_k\left[ {n\! +\! 1} \right] \right\|_{\bf{q }}^2} \big\} $ is given by
\vspace{-0.1cm}
\ifsinglecol
\vspace{-0.2cm}
\begin{align}
\E\!\left\{\! {\left\| \he_k\left[ {n\! +\! 1} \right] \right\|_{\bf{q }}^2} \right\} 
&= \E\!\left\{\! {\left\| \he\left[ nN_0\! +\! N_0 \! + \! k \right] \right\|_{\bf{q }}^2} \right\} \notag \\
&= \E\!\left\{\! {\left\| \he\left[ nN_0\! +\! N_0 \! + \! k \!-\! 1  \right] \right\|_{\F\left[((k \! - \! 1))_{N_0}\right]{\bf{q }}}^2} \right\} 
\notag \\
&\quad
 +\! 2 \mu^2\cdot \left({\bf g}^T\left[((k \! - \! 1))_{N_0}\right] \otimes \sk{((k \! - \! 1))_{N_0}}^H\right) \Pmat\left[((k \! - \! 1))_{N_0}\right] {\bf q} \notag \\
&\quad +\! \mu^2 \cdot\SigV\left[((k \! - \! 1))_{N_0}\right]\cx^T\left[((k \! - \! 1))_{N_0}\right]{\bf q}  +\! \mu^2\|{\bf g}\left[((k \! - \! 1))_{N_0}\right]\|_{\B\left[((k \! - \! 1))_{N_0}\right] {\bf q}}^2 \notag \\
&\quad -\!2\mu \cdot\!\left({\bf g}^T\!\left[((k \! - \! 1))_{N_0}\right] \! \otimes \! \tilde{\bf b}^H\!\left[ nN_0\! +\! N_0 \! + \! k \!-\! 1  \right]\right) \Pmat\left[((k \! - \! 1))_{N_0}\right] {\bf q} \notag \\
& \stackrel{(a)}{=} \E\!\left\{\! {\left\| \he\left[ nN_0\! +\! N_0 \! + \! k \!-\! 1  \right] \right\|_{\F\left[((k \! - \! 1))_{N_0}\right]{\bf{q }}}^2} \right\} 
\notag \\
&\quad 
+\! \mu^2 \pk{((k \! - \! 1))_{N_0}}^T{\bf q} -2\mu \tilde{\bf p}^T\left[ nN_0\! +\! N_0 \! + \! k \!-\! 1  \right]{\bf q},
\vspace{-0.2cm}
\label{eqn:MeanSqRec}
\end{align}
\else 
\begin{align}
&\E\!\left\{\! {\left\| \he_k\left[ {n\! +\! 1} \right] \right\|_{\bf{q }}^2} \right\} 
= \E\!\left\{\! {\left\| \he\left[ nN_0\! +\! N_0 \! + \! k \right] \right\|_{\bf{q }}^2} \right\} \notag \\
&= \E\!\left\{\! {\left\| \he\left[ nN_0\! +\! N_0 \! + \! k \!-\! 1  \right] \right\|_{\F\left[((k \! - \! 1))_{N_0}\right]{\bf{q }}}^2} \right\} \notag \\
&\quad +\! 2 \mu^2 \left({\bf g}^T\left[((k \! - \! 1))_{N_0}\right] \otimes \sk{((k \! - \! 1))_{N_0}}^H\right) \Pmat\left[((k \! - \! 1))_{N_0}\right] {\bf q} \notag \\
&\quad +\! \mu^2 \SigV\left[((k \! - \! 1))_{N_0}\right]\cx^T\left[((k \! - \! 1))_{N_0}\right]{\bf q} \notag \\
&\quad +\! \mu^2\|{\bf g}\left[((k \! - \! 1))_{N_0}\right]\|_{\B\left[((k \! - \! 1))_{N_0}\right] {\bf q}}^2 \notag \\
&\quad -\!2\mu \!\left({\bf g}^T\!\left[((k \! - \! 1))_{N_0}\right] \! \otimes \! \tilde{\bf b}^H\!\left[ nN_0\! +\! N_0 \! + \! k \!-\! 1  \right]\right)\notag \\
&\quad \qquad \times \Pmat\left[((k \! - \! 1))_{N_0}\right] {\bf q} \notag \\
& \stackrel{(a)}{=} \E\!\left\{\! {\left\| \he\left[ nN_0\! +\! N_0 \! + \! k \!-\! 1  \right] \right\|_{\F\left[((k \! - \! 1))_{N_0}\right]{\bf{q }}}^2} \right\} \notag \\
&\quad +\! \mu^2 \pk{((k \! - \! 1))_{N_0}}^T{\bf q} -2\mu\tilde{\bf p}^T\left[ nN_0\! +\! N_0 \! + \! k \!-\! 1  \right]{\bf q},
\label{eqn:MeanSqRec}
\end{align}
\fi 
where $(a)$ follows by plugging \eqref{eqn:pkDef}, recalling that 
$\left\| {{\bf{g}}\left[ n \right]} \right\|_{\B\left[ n \right]{\bf{q}}}^2 \!=\! \left( {{{\bf{g}}^T}\left[ n \right] \otimes {{\bf{g}}^H}\left[ n \right]} \right)\!\B\left[ n \right]{\bf{q}}$. 
Using \eqref{eqn:MeanSqRec}, we next show that if \eqref{eqn:MeanSqConv} is satisfied, then $\E\!\big\{\! {\left\| \he_k\left[n \right] \right\|_{\bf{q }}^2} \big\} $ converges to a fixed and finite value for $n \rightarrow \infty$, $\forall k \in \mathcal{N}_0$. 
To that aim, define:
\vspace{-0.2cm}
	\begin{equation}
		\label{eqn:ConvDefAk}
		{\bf a}_k \triangleq \bigg( \sum\limits_{l=k}^{N_0-1+k}\pk{((l))_{N_0}}^T\FProd{l+1}{k}\bigg)^T\!\!;  \qquad
		{\bf b}_k[n] \triangleq 2\bigg(\sum\limits_{l=k}^{N_0-1+k}\tilde{\bf p}^T[nN_0+l]\FProd{l+1}{k}\bigg)^T\!\!.	
		\vspace{-0.2cm}	
	\end{equation}
%
Again, since the entries of $\Cx[k]$ are bounded  and from \ref{itm:assm2} it follows that  ${\bf{a}}_k$ and ${\bf{b}}_k\left[ n \right]$ are bounded $\forall k \in \mathcal{N}_0$. 
Using these definitions, repeating \eqref{eqn:MeanSqRec} $N_0$ times results in
\vspace{-0.2cm}
\ifsinglecol
\begin{equation}
\E\left\{ {\left\|  \he_k\left[ {n \! + \! 1} \right] \right\|_{\bf{q}}^2} \right\} 
\!=\! \E\left\{ {\left\|  \he_k \left[ n \right] \right\|_{\FProd{k}{k}{\bf q}}^2} \right\}  \!+\!  {\mu ^2} \cdot {{\bf{a}}_k^T}{\bf{q}} \! -\!  2\mu \cdot {{\bf{b}}_k^T}\left[ n \right]{\bf{q}}. 
\label{eqn:VarRelationMod1} 
\vspace{-0.2cm}
\end{equation} 
\else
\begin{align}
\E\left\{ {\left\|  \he_k\left[ {n \! + \! 1} \right] \right\|_{\bf{q}}^2} \right\} 
\!&=\! \E\left\{ {\left\|  \he_k \left[ n \right] \right\|_{\FProd{k}{k}{\bf q}}^2} \right\}  \!+\!  {\mu ^2} \cdot {{\bf{a}}_k^T}{\bf{q}} \notag \\
&\quad \! -\!  2\mu \cdot {{\bf{b}}_k^T}\left[ n \right]{\bf{q}}. 
\label{eqn:VarRelationMod1} 
\end{align} 
\fi 
%
%
Following  \cite[Ch. 24.2]{Sayed:08} and \cite[Appendix F]{Shlezinger:16}, we use \eqref{eqn:VarRelationMod1} to formulate $M^2$ state-space recursions for each $k \in \mathcal{N}_0$ as follows:
\vspace{-0.2cm}
\ifsinglecol
\begin{equation}
\E\left\{ {\left\| \he_k\left[ {n \!+ \! 1} \right] \right\|_{{\left(\FProd{k}{k}\right)^{l}}{\bf{q}}}^2} \right\} \!= \! \E\left\{ {\left\| \he_k\left[ n \right] \right\|_{{\left(\FProd{k}{k}\right)^{l\!+\!1}}{\bf{q}}}^2} \right\}   \! + \! {\mu ^2}{{\bf{a}}_k^T}{\left(\FProd{k}{k}\right)^{l}}{\bf{q}}\! - \! 2\mu {{{\bf{b}}_k^T}\left[ n \right]{\left(\FProd{k}{k}\right)^{l}}{\bf{q}}},
\label{eqn:VarRelationMod1b} 
\vspace{-0.21cm}
\end{equation}
\else 
\begin{align}
\E\left\{ {\left\| \he_k\left[ {n \!+ \! 1} \right] \right\|_{{\left(\FProd{k}{k}\right)^{l}}{\bf{q}}}^2} \right\} &\!= \! \E\left\{ {\left\| \he_k\left[ n \right] \right\|_{{\left(\FProd{k}{k}\right)^{l\!+\!1}}{\bf{q}}}^2} \right\}  \notag \\
&\quad \! + \! {\mu ^2}{{\bf{a}}_k^T}{\left(\FProd{k}{k}\right)^{l}}{\bf{q}}\notag \\
&\quad\! - \! 2\mu {{{\bf{b}}_k^T}\left[ n \right]{\left(\FProd{k}{k}\right)^{l}}{\bf{q}}},
\label{eqn:VarRelationMod1b} 
\end{align}
\fi 
$l \in \{0,1,\ldots,M^2-1\}$. 
Let $\left\{\alpha_l\right\}_{l=0}^{M^2-1}$ be the coefficients of the characteristic polynomial of $\FProd{k}{k}$ \cite[Pg. 492]{Meyer:00}. It follows from the Cayley-Hamilton theorem \cite[Pg. 532]{Meyer:00} and  the linearity of the weighted  norm \cite[Eq. (23.31)]{Sayed:08} that 
\vspace{-0.1cm}
\ifsinglecol
\begin{equation*}
\E\!\Big\{\! \left\| \he_k\!\left[ n \right] \right\|_{\left( \FProd{k}{k}\right)^{{M^2}}{\bf{q}}}^2 \Big\}\! =\!  -\! \sum\limits_{l = 0}^{{M^2}\! -\! 1}\! {\alpha _l}\E\!\Big\{\! \left\| \he_k\!\left[ n \right] \right\|_{\left( \FProd{k}{k}\right)^{l}{\bf{q}}}^2 \Big\}. 
\vspace{-0.1cm}
\end{equation*}
\else 
$\E\!\left\{\! \left\| \he_k\!\left[ n \right] \right\|_{\left( \FProd{k}{k}\right)^{{M^2}}{\bf{q}}}^2 \right\}\! =\!  -\! \sum\limits_{l = 0}^{{M^2}\! -\! 1}\! {\alpha _l}\E\!\left\{\! \left\| \he_k\!\left[ n \right] \right\|_{\left( \FProd{k}{k}\right)^{l}{\bf{q}}}^2 \right\}$. 
\fi 
Thus, by defining the $M^2\times 1$ vectors $\utilde{\bf{\bar h}}_k\left[ n \right]$, $\utilde{\bf{a}}_k$, and $\utilde{\bf{b}}_k\left[ n \right]$ via 
$\left(\utilde{\bf{\bar h}}_k\!\left[ n \right]\right)_l \!\triangleq\! \E\!\left\{\! \left\| \he_k\!\left[ n \right] \right\|_{{\left( \FProd{k}{k}\right)^l}{\bf{q}}}^2 \right\}$, $\left(\utilde{\bf{a}}_k\right)_l \!\triangleq\! {\bf{a}}_k^T{\left( \FProd{k}{k}\right)^l}{\bf{q}}$, and $\left(\utilde{\bf{b}}_k\!\left[ n \right]\right)_l \triangleq {\bf{b}}_k^T\!\left[ n \right]{\left( \FProd{k}{k}\right) ^l}{\bf{q}}$,
 $l \in \{0,1,\ldots,M^2\!-\!1\}$, and the $M^2\times M^2$ matrix $\utilde{\bar{\F}}_k$ s.t. $\utilde{\bar{\F}}_k^T$ is the companion matrix of the characteristic polynomial of $\FProd{k}{k}$ \cite[Pg. 648]{Meyer:00}, 
the state-space recursions \eqref{eqn:VarRelationMod1b} can be written as a set of $N_0$ multivariate difference equations 
\vspace{-0.2cm}
\begin{equation}
\label{eqn:StateSpace1}
\utilde{\bf{\bar h}}_k\left[ {n + 1} \right] = \utilde{\bar{\F}}_k\utilde{\bf{\bar h}}_k\left[ n \right] + {\mu ^2}\utilde{\bf{a}}_k - 2\mu  {\utilde{\bf{b}}_k\left[ n \right]}, \quad k \in \mathcal{N}_0, n \geq 0. 
\vspace{-0.2cm}
\end{equation}
Note that $\forall k \in \mathcal{N}_0$, Eq. \eqref{eqn:StateSpace1} represents an $M^2\times M^2$ multivariate \ac{lti} system with input signal ${\mu ^2}\utilde{\bf{a}}_k\! -\! 2\mu  {\utilde{\bf{b}}_k\!\left[ n \right]}$ and output signal $\utilde{\bf{\bar h}}_k\!\left[ n \right]$. 
Since the entries of  $\tilde{\bf p}[n]$ are bounded $\forall n \in \mathds{N}$ and $\mathop {\lim }\limits_{n \to \infty }\tilde{\bf p}[n]\! = \! {\bf 0}_{M^2 \times 1}$, it follows that the entries of ${\utilde{\bf{b}}_k\!\left[ n \right]}$ are also bounded $\forall n \in \mathds{N}$ and that 
$\mathop {\lim }\limits_{n \to \infty }{\utilde{\bf{b}}_k\!\left[ n \right]}\! = \! {\bf 0}_{M^2 \times 1}$. 
It therefore follows from \cite[Ch. 23.4]{Sayed:08} that  $\utilde{\bf{\bar h}}_k\!\left[ n \right]$ is bounded and tends to a steady-state value for $n\! \rightarrow\! \infty$, i.e., $\E\!\left\{ \left\| \he_k\left[ n \right] \right\|_{\bf q}^2 \right\}$ is convergent, if and only if  $\rho\left( \utilde{\bar{\F}}_k\right) < 1$. 

So far we have shown that when the \ac{lms} filter is mean convergent and the entries of $\Cx[k]$ are bounded $\forall k \in \mathds{N}_0$, then it also mean-square stable if and only if  $\rho\left( \utilde{\bar{\F}}_k\right) < 1$ for all $k \in \mathds{N}_0$. Note that it follows from \cite[Pg. 346]{Sayed:08} that the eigenvalues of $\utilde{\bar{\F}}_k$  are the eigenvalues of $\FProd{k}{k}$. Therefore, when the entries of $\Cx[k]$ are bounded $\forall k \in \mathds{N}_0$, a mean convergent \ac{lms} filter is also mean-square stable if and only if $\forall k \in \mathcal{N}_0$, $\rho\left(\FProd{k}{k}\right) < 1$. 
\qed

\vspace{-0.2cm}
\subsection{Proof of Corollary \ref{cor:MeanSqConv}}
\label{app:proofMeanSqConv2}
\vspace{-0.2cm}
To prove the corollary, we show that \eqref{eqn:MeanSqCond1} is satisfied when the constraints on $\A[k]^{-1}\B[k]$ and $\Hmat[k]$ in \eqref{eqn:MeanSqConv} are satisfied.
As in Appendix \ref{app:ProofCorMeanCond}, it can be shown that $\forall k \in \mathcal{N}_0$, $\rho\left(\FProd{k}{k}\right) \leq \prod\limits_{l=0}^{N_0-1} \rho\left(\F[l] \right)$. Therefore, if $\rho\left(\F[k] \right) < 1$ for all $k \in \mathcal{N}_0$, then mean-square stability is obtained. 
Note that $\F[k]$ is Hermitian, thus its eigenvalues are real, and $\rho\left(\F[k] \right) < 1$ if and only if all the eigenvalues of $\F[k]$ are in the $(-1,1)$ interval. 
Also note that $\F[k] = {\bf I}_{M^2}-\mu\A[k]+\mu^2\B[k]$, where $\A$ and $\B$ are positive semi-definite, and $\mu > 0$. 
Note that all the eigenvalues of $\F[k]$ are smaller than $1$ if and only if ${\bf I}_{M^2}- \F[k]$ is positive definite, or equivalently, $\A[k]-\mu\B[k]$ is positive definite. This is obtained when $\A[k]$ is positive definite, which is obtained when $\Cx[k]$ is positive definite, and the step-size satisfies $\mu < \frac{1}{\MaxEig{\A^{-1}[k]\B[k]}}$ \cite[Appendix A]{AlNaffouri:04}. 
Next, note that all the eigenvalues of $\F[k]$ are larger than $-1$ if and only if $2{\bf I}_{M^2}-\mu\A[k]+\mu^2\B[k]$ is positive definite. It follows from \cite[Appendix A]{AlNaffouri:04} that this is satisfied for all $\mu \!>\! 0$ when $\Hmat[k]$ has no real positive eigenvalues, and for $\mu \!< \!\frac{1}{\MaxEig{\Hmat[k]}}$ when $\Hmat[k]$ has at least one real positive eigenvalue. 
Combining the conditions used in the proof yields \eqref{eqn:MeanSqConv}. 
\qed  

\vspace{-0.2cm}
\subsection{Proof of Theorem \ref{thm:ssMSE}}
\label{app:proofSteady}
\vspace{-0.2cm}
For $k \in \mathcal{N}_0$, $n \in \mathds{N}$, using definition \eqref{eqn:hkDef} and the periodicity of ${\bf g}[n]$, $\Cx[n]$, and $\SigV[n]$, the instantaneous \ac{mse} \eqref{eqn:ETaMSECurve} can be written as 
\vspace{-0.2cm}
\ifsinglecol
\begin{equation}
\E\left\{|e\left[ n\cdot N_0 \!+\! k \right]|^2\right\} 
\!=\! \E\left\{ \left\| \he_k\left[ n \right] \right\|_{\cx[k]}^2 \right\}   \!+\!2 {\rm Re}\left\{{\bf g}^H[k]\Cx[k]\E\{\he_k[n]\}\right\}
\!+\!{\left\| {\bf g}\left[ k \right] \right\|_{\cx[k]}^2} 
\!+\! \SigV[k].
\label{eqn:ssMSE1}
\vspace{-0.2cm}
\end{equation}
\else
\begin{align}
\E\left\{|e\left[ n\cdot N_0 + k \right]|^2\right\} 
&= \E\left\{ \left\| \he_k\left[ n \right] \right\|_{\cx[k]}^2 \right\}  \notag \\
&\quad +2 {\rm Re}\left\{{\bf g}^H[k]\Cx[k]\E\{\he_k[n]\}\right\}
\notag \\
&\quad +{\left\| {\bf g}\left[ k \right] \right\|_{\cx[k]}^2} 
+ \SigV[k].
\label{eqn:ssMSE1}
\vspace{-0.1cm}
\end{align}
\fi 
When Thm.  \ref{thm:MeanSqConv} is satisfied, then  the adaptive filter is mean convergent and mean-square stable as in Defs. \ref{def:MeanConv1}--\ref{def:Stable}, respectively. Thus, letting $n \rightarrow \infty$ in \eqref{eqn:ssMSE1} yields
\vspace{-0.2cm}
\ifsinglecol
\begin{align}
\mathop {\lim }\limits_{n \to \infty }\E\left\{|e\left[ n\cdot N_0 \! + \! k \right]|^2\right\}  
&=\mathop {\lim }\limits_{n \to \infty }  \E\left\{ \left\| \he_k\left[ n \right] \right\|_{\cx[k]}^2 \right\}   \! + \!2 {\rm Re}\!\left\{\!{\bf g}^H\![k]\Cx[k]\!\mathop {\lim }\limits_{n \to \infty }\!\E\{\he_k[n]\}\right\}\notag \\
&\quad \! + \!{\left\| {\bf g}\left[ k \right] \right\|_{\cx[k]}^2} 
\! + \! \SigV[k].
\label{eqn:ssMSE2}
\vspace{-0.2cm}
\end{align}
\else
\begin{align}
\mathop {\lim }\limits_{n \to \infty }\E\left\{|e\left[ n\cdot N_0 \! + \! k \right]|^2\right\}  
&=\mathop {\lim }\limits_{n \to \infty }  \E\left\{ \left\| \he_k\left[ n \right] \right\|_{\cx[k]}^2 \right\}  \notag \\
& \! + \!2 {\rm Re}\!\left\{\!{\bf g}^H\![k]\Cx[k]\!\mathop {\lim }\limits_{n \to \infty }\!\E\{\he_k[n]\}\right\}\notag \\
& \! + \!{\left\| {\bf g}\left[ k \right] \right\|_{\cx[k]}^2} 
\! + \! \SigV[k].
\label{eqn:ssMSE2}
\vspace{-0.1cm}
\end{align}
\fi 
Next, recalling the definitions of ${\bf a}_k$ and  ${\bf b}_k[n]$ stated in \eqref{eqn:ConvDefAk}, it follows from \eqref{eqn:VarRelationMod1} that $\forall k \in \mathcal{N}_0$, for $n \rightarrow \infty$ it holds that
\ifsinglecol
\vspace{-0.2cm}
\begin{align*}
\!\mathop {\lim }\limits_{n \to \infty }\! \E\!\left\{ {\left\| \he_k \left[ n \right] \right\|_{{\bf{q}}}^2} \right\}\! 
&
= \! \mathop {\lim }\limits_{n \to \infty }\! \E\!\left\{ \left\| \he_k \left[ n \right] \right\|_{\FProd{k}{k}{\bf{q}}}^2 \right\}\! +\! {\mu ^2}\cdot{\bf a}_k^T{\bf{q}}  -2\mu \!\cdot\! \mathop {\lim }\limits_{n \to \infty }{{{\bf{b}}_k^T}\left[ n \right]{\bf{q}}} 
\notag \\
&
\stackrel{(a)}{=} \!\mathop {\lim }\limits_{n \to \infty }\! \E\!\left\{ \left\| \he_k \left[ n \right] \right\|_{\FProd{k}{k}{\bf{q}}}^2 \right\}\! +\! {\mu ^2}\cdot{\bf a}_k^T{\bf{q}},
\vspace{-0.2cm}
\end{align*}
\else
\begin{align*}
\!\mathop {\lim }\limits_{n \to \infty }\! \E\!\left\{ {\left\| \he_k \left[ n \right] \right\|_{{\bf{q}}}^2} \right\}\! 
&= \! \mathop {\lim }\limits_{n \to \infty }\! \E\!\left\{ \left\| \he_k \left[ n \right] \right\|_{\FProd{k}{k}{\bf{q}}}^2 \right\}\! +\! {\mu ^2}{\bf a}_k^T{\bf{q}} \notag \\
&\quad -2\mu \!\cdot\! \mathop {\lim }\limits_{n \to \infty }{{{\bf{b}}_k^T}\left[ n \right]{\bf{q}}} \notag \\
&\stackrel{(a)}{=} \!\mathop {\lim }\limits_{n \to \infty }\! \E\!\left\{ \left\| \he_k \left[ n \right] \right\|_{\FProd{k}{k}{\bf{q}}}^2 \right\}\! +\! {\mu ^2}{\bf a}_k^T{\bf{q}},
\end{align*}
\fi 
where $(a)$ follows since  $\mathop {\lim }\limits_{n \to \infty } {\bf b}_k[n] = {\bf 0}_{M \times 1}$.
Thm.  \ref{thm:MeanSqConv} guarantees that both  $\mathop {\lim }\limits_{n \to \infty } \E\big\{ {\left\| \he_k \left[ n \right] \right\|_{{\bf{q}}}^2} \big\}$ and $\mathop {\lim }\limits_{n \to \infty } \E\big\{ {\left\| \he_k \left[n \right] \right\|_{\FProd{k}{k}{\bf{q}}}^2} \big\}$   exist and are finite. Thus,  from the linearity of the weighted Euclidean norm \cite[Eq. (23.31)]{Sayed:08} we have that
\vspace{-0.2cm}
\begin{equation}
\label{eqn:SteadyVar2}
\mathop {\lim }\limits_{n \to \infty } \E\big\{ {\left\| \he_k \left[ n \right] \right\|_{\left( {\bf{I}}_{{M^2}} - \FProd{k}{k} \right){\bf{q}}}^2} \big\}  = {\mu ^2}{\bf a}_k^T{\bf{q}}, \qquad \forall k \in \mathcal{N}_0.
\vspace{-0.2cm}
\end{equation}
Setting ${\bf q} = \left( {\bf{I}}_{M^2} - \FProd{k}{k} \right)^{ - 1}\cx [k]$ in \eqref{eqn:SteadyVar2}
 yields
%
\ifsinglecol
\vspace{-0.2cm}
\begin{align}
\mathop {\lim }\limits_{n \to \infty }\! \E\!\left\{\!\left\| \he_k\left[ n \right] \right\|_{\cx [k]}^2 \right\}\! 
&= \!{\mu ^2}{\bf a}_k^T \! \left( {\bf{I}}_{M^2} \!-\! \FProd{k}{k} \right)^{\! -\! 1}\!\cx [k] \notag \\
& \stackrel{(a)}{=} \mu^2\!\!\sum\limits_{l=k}^{N_0\!+\!k\!-\!1}\!\!\pk{((l))_{N_0}}^T\FProd{l\!+\!1}{k}\left({\bf I}_{M^2}\!-\!\FProd{k}{k}\right)^{-1}\cx[k],
\vspace{-0.4cm}
\label{eqn:SteadyVar2b}
\end{align}
\else
\begin{align}
&\!\!\!\mathop {\lim }\limits_{n \to \infty }\! \E\!\left\{\!\left\| \he_k\left[ n \right] \right\|_{\cx [k]}^2 \right\}\! 
= \!{\mu ^2}{\bf a}_k^T \! \left( {\bf{I}}_{M^2} \!-\! \FProd{k}{k} \right)^{\! -\! 1}\!\cx [k] \notag \\
&\qquad \stackrel{(a)}{=} \mu^2\!\!\sum\limits_{l=k}^{N_0\!+\!k\!-\!1}\!\!\pk{((l))_{N_0}}^T\FProd{l\!+\!1}{k}\left({\bf I}_{M^2}\!-\!\FProd{k}{k}\right)^{-1}\cx[k],
\label{eqn:SteadyVar2b}
\end{align}
\fi 
where $(a)$ follows from \eqref{eqn:ConvDefAk}. Plugging \eqref{eqn:SteadyVar2b} and \eqref{eqn:MeanSSHka} into \eqref{eqn:ssMSE2}, noting that $\sk{k} = \big({\bf I}_{M} \! - \! \RxProd{k}{k} \big)^{\! - \!1}\bk{k}$ in \eqref{eqn:MeanSSHka}, yields \eqref{eqn:ssMSE}. 
\qed 


\end{appendix}


\end{document}